\documentclass[aps,pre,twocolumn,showpacs,amsfonts]{revtex4}
\usepackage{epsfig}
\usepackage{graphicx}
\usepackage{dcolumn}
\usepackage{bm}

\begin{document}

\title{A nonlinear theory of non-stationary low Mach number channel flows of freely cooling nearly elastic granular gases}

\author{Baruch Meerson}
\author{Itzhak Fouxon}
\author{Arkady Vilenkin}

\affiliation{Racah Institute of Physics, Hebrew University of Jerusalem,
Jerusalem 91904, Israel}
\date{\today }

\begin{abstract}
We employ hydrodynamic equations to investigate non-stationary channel flows of
freely cooling dilute gases of hard and smooth spheres with nearly elastic
particle collisions. This work focuses on the regime where
the sound travel time through the channel is much shorter than the
characteristic cooling time of the gas. As a result, the gas
pressure rapidly becomes almost homogeneous, while the typical Mach number of
the flow drops well below unity. Eliminating the acoustic modes and employing
Lagrangian coordinates, we reduce the hydrodynamic equations to a single
nonlinear and nonlocal equation of a reaction-diffusion type. This equation
describes a broad class of channel flows and, in particular, can follow the development
of the clustering instability from a weakly perturbed homogeneous cooling state
to strongly nonlinear states. If the heat diffusion is neglected, the reduced
equation becomes exactly soluble, and the solution develops a finite-time
density blowup. The blowup has the same local features at singularity as those
exhibited by the recently found family of exact solutions of the full set of
ideal hydrodynamic equations (Fouxon \textit{et al.} 2007). The heat diffusion,
however, always becomes important near the attempted singularity. It arrests the
density blowup and brings about novel inhomogeneous cooling states (ICSs) of
the gas, where the pressure continues to decay with time, while the density
profile becomes time-independent. The ICSs  represent exact solutions of the
full set of granular hydrodynamic equations. Both the density profile of an ICS,
and the characteristic relaxation time towards it are determined by a single
dimensionless parameter ${\cal L}$ that describes the relative role of the
inelastic energy loss and heat diffusion.  At ${\cal L} \gg 1$ the intermediate cooling
dynamics proceeds as a competition between ``holes": low-density regions of the
gas. This competition resembles Ostwald ripening (only one hole survives at the
end), and we report a particular regime where the ``hole ripening" statistics exhibits a
simple dynamic scaling behavior.

\end{abstract}


\maketitle
\section{Introduction}

Clustering of matter is a spectacular example of structure formation in nature.
A fascinating example of clustering is provided by granular gases: gases of
macroscopic particles that lose kinetic energy in collisions. Granular gas is a
low-density limit of granular flows \cite{BP,Goldhirsch2}. The simplest version
of the granular gas model assumes a dilute assembly of identical smooth hard
spheres (with diameter $\sigma$ and unit mass) who lose energy at binary
collisions in such a way that the normal component of the relative velocity of
the colliding particles gets reduced by a constant factor $0\leq r<1$ (the
coefficient of normal restitution) upon each collision. Granular gases exhibit
various pattern forming instabilities, including the shearing/clustering instability of a
freely cooling homogeneous inelastic gas
\cite{Hopkins,Goldhirsch,McNamara1,McNamara2,Ernst,Brey,Luding,van
Noije,Ben-Naim2,ELM,MP,Garzo}. This instability causes the generation of a
macroscopic flow, both solenoidal and potential, and formation of dense clusters of particles.

A natural theoretical description of macroscopic granular flows is provided by
the Navier-Stokes granular hydrodynamics \cite{BP,Goldhirsch2}. Although the
criteria of its validity are quite restrictive, see below, granular
hydrodynamics has a great predictive power, sometimes going far beyond the
formal limits of applicability \cite{Goldhirsch2}. Recently, granular
hydrodynamics has been applied to a variety of \textit{non-stationary} flows of
granular gases \cite{ELM,Bromberg,Volfson,Fouxon1,Fouxon2}. Non-stationary flows
provide sharp tests to continuum models of granular flows, especially when the
time-dependent solutions of the continuum equations tend to develop finite-time
singularities. Examples are provided by the recently predicted finite-time
blowup of the gas density in freely cooling granular gases: at zero gravity
\cite{ELM,Fouxon1,Fouxon2} (as described by ideal granular hydrodynamic
equations), and at finite gravity (even in the framework of non-ideal granular hydrodynamic
equations) \cite{Volfson}.

We will assume in this paper that particle collisions are almost
elastic,  the local gas density (that we denote by $\rho$) is
\textit{much} smaller than the close-packing density, and the
Knudsen number is very small:
\begin{equation}\label{threeineq}
    1-r\ll 1\,,\;\;\;\;\;\rho \sigma^d\ll 1\,,\;\;\;\;
    \mbox{and}\;\;\;\;\;l_{free}/L\ll1\,.
\end{equation}
Here $d>1$ is the dimension of space, $l_{free}$ is the mean free path of the
particles, and $L$ is the characteristic length scale of the hydrodynamic
fields. Under these assumptions (the second and third ones need to be verified
\textit{a posteriori}) the Navier-Stokes hydrodynamics provides a quantitatively
accurate leading-order theory \cite{BP,Goldhirsch2}. It was shown
\cite{Goldhirsch,McNamara1}, by using hydrodynamic equations that, for
sufficiently large systems, the homogeneous cooling state (HCS) of the granular
gas becomes unstable with respect to small perturbations. There are two linearly
unstable modes. The shear mode corresponds to the development of a macroscopic
solenoidal flow, while the clustering mode corresponds to the development of a
macroscopic potential flow that brings about formation of clusters of particles.

A consistent \textit{nonlinear} hydrodynamic theory of the clustering instability has not been available
for quite a long time. Solving strongly nonlinear hydrodynamic equations is hard (even numerically), and one looks for additional simplifications. Following Refs. \cite{ELM,MP,Fouxon1,Fouxon2}, we will assume throughout this paper that the macroscopic flow (but not microscopic motion of the particles!) is one-dimensional (1d). This assumption is natural in the geometry of a narrow channel with perfectly
elastic side walls that we adopt
here. In a narrow channel both the
clustering mode in the transverse directions, and the shear mode are suppressed (see Refs.
\onlinecite{ELM,MP} for detail).  As a result, the
macroscopic flow can depend only on the coordinate along
the channel and time, and we can focus on the development
of the pure clustering mode as it enters a strongly nonlinear
regime.

Efrati \textit{et al.} \cite{ELM} investigated numerically the long-wavelength
limit of such a quasi-1d clustering instability. In this limit the inelastic energy
loss of the gas is the fastest process, so the gas pressure rapidly drops to a
very small value. The further dynamics becomes (almost) purely inertial which
(almost) brings about a finite-time blow-up of the velocity gradient and,
therefore, of the density \cite{Whitham}. The signatures of this finite-time
singularity were indeed observed in the numerical solution of the hydrodynamic
equations \cite{ELM} until the growing gas density became so high that the
numerical scheme lost accuracy. The numerical results of Ref. \cite{ELM} were
tested in molecular dynamics (MD) simulations \cite{MP}. The MD simulations
supported the free-flow blow-up scenario until the time when the gas density
approached the hexagonal close-packing value, and the further density growth
stopped.

Recently, Fouxon et al. \cite{Fouxon1,Fouxon2} analyzed, analytically and numerically, the one dimensional flow in the framework of equations of \textit{ideal} hydrodynamics
(that is, neglecting the heat diffusion and viscosity effects). They derived a family of exact solutions to these equations, with and without shocks, for which an initially smooth flow develops a finite-time density blowup. Close to the blow-up time $t_c$, the maximum density exhibits a power law behavior $\sim
(t_c-t)^{-2}$. The velocity gradient blows up as $\sim  - (t_c-t)^{-1}$, whereas
the velocity itself remains continuous and develops a cusp, rather than a shock
discontinuity, at the singularity. The gas temperature vanishes at the
singularity, but the pressure remains finite. Extensive numerical simulations with the ideal hydrodynamic equations showed that the singularity exhibited by the exact solutions is universal, as it develops for generic initial conditions. Very recently, the existence of the attempted blowup regime has been proved in molecular dynamic simulations of a gas of nearly elastically colliding hard disks in a channel geometry \cite{Puglisi}. The results of Refs. \cite{Fouxon1,Fouxon2} also imply that, for long wavelength initial conditions, the free flow regime may not hold all the way to the density blowup \cite{Fouxon1,Fouxon2}. Very close to the
attempted free-flow singularity, compressional heating starts to act. As a
result, the gas pressure again becomes important and changes the
local blowup properties.

A crucial feature of the finite-time singularity
of the ideal hydrodynamic equations
is that it obeys an isobaric scenario: the (finite) gas pressure becomes
uniform in space in a close vicinity of the
developing singularity \cite{Fouxon2}. This hints at the possibility of an additional simplification of the problem. Indeed, an (almost) homogeneous pressure in a gas implies a low Mach number flow, when the inertial terms in the momentum equation are small compared to the pressure gradient term. This regime
appears when the sound travel time through the system is very short compared with other time scales of the problem, and one is interested in the dynamics of the system at the long time scales \cite{Zeldovich,Meerson89a,Meerson89b,AMS1,AMS2,Kaganovich,Glasner,MeersonRMP}. In particular, this regime appears naturally in the linear theory of the clustering instability of the HCS for intermediate wavelengths of the perturbations, see below. It is this (almost) spatially independent pressure regime that we will be considering in the present work.

The remainder of the paper is organized as follows. In Section \ref{hydro} we
start with a full set of equations of granular hydrodynamic for a dilute
granular flow in a channel and reduce them, for sufficiently short channels, to the
low Mach number flow equations.  In Section \ref{Lagrange} we employ Lagrangian
coordinates which enable us to exactly reduce the low Mach number flow equations
to a single nonlinear and nonlocal equation, of a reaction-diffusion type, for
the square root of the inverse gas density.  The new equation is tested in
Section \ref{tests} on two simple problems: the HCS and the linear theory of
clustering instability in short channels. In Section \ref{blowup} we show that,
when the heat diffusion is neglected, the new equation becomes exactly soluble,
and the solution develops a finite-time density blowup with the same universal
features at singularity as those exhibited by the family of exact solutions of
the full set of ideal granular hydrodynamic equations \cite{Fouxon1,Fouxon2}.
Section \ref{arrest} presents an analytical and numerical analysis that shows
that the heat diffusion term, no matter how small in the beginning,
becomes important near the attempted density blowup. As a result, the density blowup is
arrested, and a novel, inhomogeneous cooling state (ICS) of the gas emerges, with a time-independent inhomogeneous density profile.  Importantly, the ICSs represent \textit{exact} solutions of hydrodynamic equations. A limiting form of the novel cooling state is
what we call the ``hole", and we investigate its properties and the relaxation
dynamics towards it.    For sufficiently long channels (other parameters being
fixed) the cooling dynamics of the system takes the form of a competition
between, and ``ripening" of, holes.  Therefore, in Section \ref{coarsening} we
investigate the dynamics and statistics of this competition. In Section
\ref{summary} we summarize our results and put them into a perspective.

\section{Granular hydrodynamics and a low Mach number flow}
\label{hydro}%

For flows depending on a single spatial coordinate $x$ and time $t$ the granular
hydrodynamic equations can be written as follows:
\begin{eqnarray}&&
\frac{\partial \rho}{\partial t}+\frac{\partial(\rho v)}{\partial x}=0,
\label{hydrodynamics1}\\&& \rho\left(\frac{\partial v}{\partial t}+
v\frac{\partial v}{\partial x}\right)=-\frac{\partial (\rho T)}{\partial
x}+\nu_0\frac{\partial}{\partial x}\left(\sqrt{T}\frac{\partial v}{\partial
x}\right), \label{hydrodynamics2}
\\&& \frac{\partial T}{\partial t}+
v\frac{\partial T}{\partial x}=-(\gamma-1) T\frac{\partial v}{\partial
x}-\Lambda\rho
T^{3/2}\nonumber\\&&+\frac{\kappa_0}{\rho}\frac{\partial}{\partial x}
\left(\sqrt{T}\frac{\partial T}{\partial x}\right)
+\frac{\nu_0(\gamma-1)\sqrt{T}}{\rho}\left(\frac{\partial v}{\partial
x}\right)^2\,.\label{hydrodynamics3}
\end{eqnarray}
Here $\gamma$ is the adiabatic index of the gas ($\gamma=2$ and $5/3$ for $d=2$
and $d=3$, respectively), $\Lambda=2 \pi^{(d-1)/2} (1-r^2) \sigma^{d-1}/[d\,
\Gamma(d/2)]$ (see \textit{e.g.} \cite{Brey}), $\Gamma(\dots)$ is the
gamma-function, and $d\ge2$ is the dimension of space, so that $d=2$ corresponds
to disks, and $d=3$ to hard spheres. Furthermore,
$\nu_0=(2\sigma\sqrt{\pi})^{-1}$ and $\kappa_0=4\nu_0$ in 2D, and
$\nu_0=5(3\sigma^2 \sqrt{\pi})^{-1}$ and $\kappa_0=15\nu_0/8$ in 3D \cite{BP}.
Equations~(\ref{hydrodynamics1})-(\ref{hydrodynamics3}) differ from the
hydrodynamic equations for a dilute gas of \textit{elastically} colliding
spheres only by the presence of the inelastic loss term $-\Lambda \rho T^{3/2}$
which is proportional to the average energy loss per collision, $\sim (1-r^2)T$,
and to the collision rate, $\sim \rho T^{1/2}$.

It will be convenient for our purposes to rewrite Eqs.~(\ref{hydrodynamics1})-(\ref{hydrodynamics3})
in terms of the pressure $p=\rho T$, rather than the temperature. The energy equation (\ref{hydrodynamics3}) becomes
\begin{eqnarray}&&
\frac{\partial p}{\partial t} +v \frac{\partial p}{\partial x}=-\gamma
p\frac{\partial v}{\partial x}-\Lambda\rho^{1/2} p^{3/2}\nonumber\\&&
+\kappa_0 \frac{\partial}{\partial x}
\left[\sqrt{\frac{p}{\rho}}\,\frac{\partial}{\partial x}
\left(\frac{p}{\rho}\right)\right]+\nu_0(\gamma-1)\sqrt{\frac{p}{\rho}}
\left(\frac{\partial v}{\partial x}\right)^2. \label{pressureequation}
\end{eqnarray}
A set of hydrodynamic equations can be simplified if there is a time scale separation or, equivalently, a length scale separation, in the problem. For a freely cooling granular gas, a basic time scale is the characteristic cooling time
\begin{equation}\label{tc}
t_c=\frac{2}{\Lambda\rho_0^{1/2}p_0^{1/2}}\,,
\end{equation}
where $\rho_0$ is the average gas density (the total gas
mass divided by the volume of the channel), and $p_0$ is a characteristic value of the initial pressure.
There are two characteristic length scales related to $t_c$. The first is the sound travel distance
$$
l_{s}=\frac{\gamma \sqrt{2}}{\Lambda \rho_0}\sim c_s t_c\,,
$$
which is of the order of the distance a sound wave with speed $c_s=(\gamma p_0/\rho_0)^{1/2}$ travels during the time $t_c$.  The quantity $l_s$ is the same as the length scale $l$ introduced in Refs. \cite{Fouxon1,Fouxon2}.

The second characteristic length scale is the heat diffusion length
$$
l_d \sim \left(\frac{\kappa_0 p_0^{1/2}t_c}{\rho_0^{3/2}}\right)^{1/2} \sim \frac{\kappa_0^{1/2}}{\Lambda^{1/2} \rho_0}
$$
which, up to a numerical pre-factor, coincides with the critical length
\begin{equation}
l_{cr}=\sqrt{\frac{2\kappa_0}{\Lambda \rho_0^2}}\,,
\end{equation}
predicted by the linear theory of the clustering instability. The ratio  $l_s/l_d$ is of order
$(\kappa_0 \Lambda)^{-1/2} \sim (1-r^2)^{-1/2}$. As we have already assumed a strong
inequality $1-r^2 \ll 1$, this ratio is very large: $l_s/l_d \gg 1$. Throughout the
rest of the paper we will also assume that the channel length $L$ is much shorter than the sound travel distance $l_s$. This hierarchy of length scales brings about a reduced set of equations, in much the same way as in hydrodynamics of optically thin gases and plasmas that cool by their
own radiation \cite{Zeldovich,Meerson89b,AMS1,AMS2,MeersonRMP}. Note that the length scale separation $L\ll l_s$ is equivalent to
a time scale separation: the sound travel time through the channel, $L/c_s$, is much shorter than the characteristic cooling time $t_c$. As a result, sound waves rapidly make the pressure (almost) homogeneous throughout the channel. The subsequent slower evolution of  the gas proceeds on the background of an almost homogeneous (but in general time-dependent) gas pressure, while typical Mach numbers of the flow are much less than unity. In a more formal language, this reduction of the hydrodynamic equations corresponds to elimination of acoustic modes.

Before we perform the reduction procedure, let us introduce rescaled variables. We will measure the distance along the channel
in the units of $l_{cr}$, rescale time by $t_c$, and measure the gas density, pressure and velocity in the units of $\rho_0$, $p_0$ and $l_{cr}/t_c$, respectively. Keeping the original notation for the rescaled variables, we observe that Eq.~(\ref{hydrodynamics1}) does not change, while Eqs.~(\ref{hydrodynamics2})
and (\ref{pressureequation}) become
\begin{eqnarray}&&
\varepsilon_1
\rho\left(\frac{\partial v}{\partial t}+v\frac{\partial v}{\partial x}\right)=
- \,\frac{\partial p}{\partial x}+\varepsilon_2
\frac{\partial}{\partial x}\left(\sqrt{\frac{p}{\rho}}\frac{\partial v}{\partial
x}\right), \label{a31} \\&&
\frac{\partial p}{\partial t} +v \frac{\partial p}{\partial x}=-\gamma
p\frac{\partial
v}{\partial x}-2\rho^{1/2} p^{3/2} \nonumber\\&&
+\frac{\partial}{\partial x}
\left[\sqrt{\frac{p}{\rho}}\,\frac{\partial}{\partial x}
\left(\frac{p}{\rho}\right)\right]+\varepsilon_2 \,(\gamma-1)\, \sqrt{\frac{p}{\rho}}
\left(\frac{\partial v}{\partial x}\right)^2 \,, \label{a32}
\end{eqnarray}
where $\varepsilon_1=\kappa_0 \Lambda/2$, and $\varepsilon_2=\nu_0 \Lambda/2$, and $\varepsilon_1\sim \varepsilon_2\sim 1-r^2\ll 1$. We will limit ourselves to the zeroth order approximation with respect to this small parameter and send $\varepsilon_1$ and $\varepsilon_2$ to zero.  The continuity equation (\ref{hydrodynamics1}) does not change.  The momentum equation (\ref{a31}) becomes $\partial p/\partial x = 0$, therefore $p=p(t)$ is independent of $x$. The energy equation becomes
\begin{eqnarray}
\label{simpler} \dot{p}(t)&=&-\gamma p(t)\frac{\partial
v}{\partial x}-2\rho^{1/2} p(t)^{3/2} \nonumber\\
&+& p(t)^{3/2}\frac{\partial}{\partial x}
\left[\frac{1}{\sqrt{\rho}}\,\frac{\partial}{\partial x}
\left(\frac{1}{\rho}\right)\right]\,.
\end{eqnarray}
The rescaled length of the channel is
\begin{equation}
{\cal L}=\frac{L}{l_{cr}}=
\left\{\begin{array}{ll}
 (\sqrt{\pi}/2)(1-r^2)^{1/2} \rho_0 \sigma L\;\;\;\quad \mbox{in 2d,} \\
\sqrt{16\pi/75}\,(1-r^2)^{1/2} \rho_0 \sigma^2 L \quad \mbox{in 3d}\,.
\end{array}
\right.
\label{calL}
\end{equation}
Note that, in the rescaled variables, the rescaled length of the channel ${\cal L}$ coincides with the rescaled total mass of the gas, $\int_0^{{\cal L}}\rho(x,t)\,dx$.

To get an explicit expression for $\dot{p}$ we integrate Eq.~(\ref{simpler}) over the whole channel. Assuming either periodic, or no-flux boundary conditions (BCs) at the channel ends $x=0$ and $x={\cal L}$, we obtain
\begin{equation}\label{pressure0}
   \frac{\dot{p}(t)}{p(t)^{3/2}} = -2 \left\langle \rho^{1/2}(x,t)\right\rangle_x\,,
\end{equation}
where we have introduced the spatial average
$$\left\langle \dots \right\rangle_x = \frac{1}{\cal L} \int_0^{\cal L} (\dots)\, dx\, .$$
For the low Mach number flow, Eq.~(\ref{pressure0}) describes, in the
leading order, the global energy balance of the gas, see Section \ref{arrest} C
below. Equations~(\ref{hydrodynamics1}), (\ref{simpler}) and (\ref{pressure0})
for $\rho(x,t)$, $v(x,t)$ and $p(t)$ make a complete set of reduced but fully
nonlinear equations for the low Mach number flow of a freely cooling granular
gas in a channel geometry. As is usually the case for low Mach number flows,
the viscous terms dropped from the
reduced formulation, while the heat diffusion term remains.

The rescaled length/mass of the system ${\cal L}$, see Eq.~(\ref{calL}), is
determined by the relative role of the inelastic energy loss and heat
diffusion. As we will see shortly, ${\cal L}$ controls the main properties of
the cooling dynamics. For comparison, the characteristic initial pressure $p_0$
only sets the time scale for the dynamics. To facilitate future comparisons of
the theory with MD simulations, we rewrite the parameter ${\cal L}$ in a
slightly different form:
$$
{\cal L}=
\left\{\begin{array}{ll}
\frac{\sqrt{\pi (1-r^2)}\, N_{p}\sigma}{2 L_y} \;\;\;\quad \mbox{in 2d,} \\
\frac{\sqrt{16\pi (1-r^2)}\, N_{p}\sigma^2}{\sqrt{75}
\,L_y L_z} \quad \mbox{in 3d}\,.
\end{array}
\right.
$$
Here $N_{p}$ is the total number of
particles in the channel, and $L_y$ and $L_z$ are the transverse channel dimensions.

\section{Lagrangian description and nonlocal reaction-diffusion equation}
\label{Lagrange}
Remarkably, it is possible to bring the three
equations~(\ref{hydrodynamics1}), (\ref{simpler}) and (\ref{pressure0}) to
a single nonlocal equation of a reaction-diffusion type. Let us first introduce
Lagrangian mass coordinates \cite{ZR}. It is convenient to choose a reference frame so that $v(x=0,t)=0$.  For the periodic boundary conditions (BCs)
one can always achieve this by exploiting the Galilian invariance
of the hydrodynamic equations  to get rid of the center-of-mass motion.
This sets $v(x=0,t)=0$, where $x=0$ is the center-of-mass coordinate. For the no-flux BC (impenetrable walls), a natural choice of $x=0$ is at one of the walls, where the gas
velocity is again zero. Then a convenient choice of the
Lagrangian mass coordinate is
\begin{equation}
m(x, t)=\int_0^x \rho(x^{\prime}, t)dx^{\prime}\,, \label{masscoordinate}
\end{equation}
which is simply the (rescaled) mass content between the Eulerian points $0$ and $x$.
The inverse transformation is
\begin{equation}
x(m, t)=\int_0^m \frac{dm^{\prime}}{\rho(m^{\prime}, t)}\,. \label{space}
\end{equation}
In the Lagrangian coordinates Eqs.~(\ref{hydrodynamics1}) and (\ref{simpler})  become
\begin{eqnarray}
\frac{\partial}{\partial t} \left(\frac{1}{\rho}\right)&=& \frac{\partial
v}{\partial m},\label{eqs1}\\
\dot{p}&=&-\gamma p \rho\frac{\partial v}{\partial m}-2 p^{3/2}\rho^{1/2}
\nonumber \\&+& p^{3/2} \rho \frac{\partial}{\partial
m}\left(\sqrt{\rho} \frac{\partial}{\partial m}\frac{1}{\rho}\right)\,.
\label{eqs2}
\end{eqnarray}
As the total rescaled mass of the gas is equal to the rescaled channel length ${\cal L}$, we define the spatial average in the Lagrangian coordinate as
$$\left\langle \dots \right\rangle = \frac{1}{{\cal L}} \int_0^{\cal L} (\dots)\, dm\,,$$ and
rewrite Eq.~(\ref{pressure0}) as
\begin{eqnarray}
\frac{\dot{p}(t)}{p(t)^{3/2}} &=&-2 \left\langle
\frac{1}{\rho^{1/2}(m,t)}\right\rangle\,.\label{pressure}
\end{eqnarray}
It is convenient to introduce a new rescaled variable
$w(m,t) = \rho^{-1/2}(m,t)$ and a new rescaled time
\begin{equation}\label{tau}
\tau=\frac{1}{\gamma}
\int_0^t p^{1/2}(t') dt'\,.
\end{equation}
Then, by eliminating $\partial_m v$ and $\dot{p}$ from
Eqs.~(\ref{eqs1})-(\ref{pressure}), we can reduce these equations to a single
integro-differential equation of a reaction-diffusion type:
\begin{equation} \label{eq5}
 w \frac{\partial w}{\partial \tau} =
 -w+w^2\,\left\langle w\right\rangle+\, \frac{\partial^2 w}{\partial m^2}\,.
\end{equation}
Equation~(\ref{eq5}) describes a broad class of slow 1d flows in freely cooling nearly elastic granular gases. In particular, this equation encodes the development of the
clustering instability: from a weakly perturbed HCS (after a brief acoustic transient) all the way to the strongly nonlinear stage. Indeed, let us
rewrite Eq.~(\ref{pressure}) in terms of the new variable $w(m,\tau)$ and
new time $\tau$:
\begin{equation}\label{pressure1}
    \frac{1}{p(\tau)} \frac{dp}{d\tau}=-2 \gamma \left\langle w(m,\tau)\right\rangle\,.
\end{equation}
Once Eq.~(\ref{eq5}) for  $w(m,\tau)$  is solved, we can calculate
the pressure $p(\tau)$ from Eq.~(\ref{pressure1}) and then return to the
(rescaled) physical time $t$ using Eq.~(\ref{tau}):
\begin{equation}\label{t}
    t=\gamma \int_0^{\tau} \frac{d\tau'}{p^{1/2}(\tau')}\,.
\end{equation}
Furthermore, using Eq.~(\ref{eqs1}) and the condition $v(m=0, t)=0$, we can find
the gas velocity: $v(m, t)=\int_0^m \partial_t w^2(m^{\prime}, t)dm^{\prime}$. Finally, we can return to
the Eulerian coordinate by using Eq.~(\ref{space}): $x(m, t)=\int_0^m w^2(m^{\prime}, t)dm^{\prime}$.

Notably, equation~(\ref{eq5}) is parameter-free: the only parameter entering the problem [except possible parameters introduced by the initial condition $w(m,0)$] is the  rescaled system length/mass ${\cal L}$.  Conservation of the total mass of the gas in the channel appears in the Lagrangian formulation as the conservation law
\begin{equation}\label{conslaw}
    \langle w^2(m,\tau)\rangle=1\,,
\end{equation}
easily verifiable from Eq.~(\ref{eq5}).

\section{Simple tests: HCS and linear theory of clustering instability}
\label{tests}

As a first test of Eqs.~(\ref{eq5}) and (\ref{pressure1}), let us consider a
HCS. Here at $t=0$ we have (in the physical units)
$\rho(m,t=0)=\rho_0=const$,
$T(m,t=0)=T_0=const$ and $v(m,t=0)=0$ and, therefore, $w(m,t=0)= 1$ and
$p(t=0)=p_0=\rho_0 T_0$. As the gas density remains constant in space at $t>0$, we can
rewrite Eq.~(\ref{eq5}) as
\begin{equation} \label{uniform}
  \frac{dw(\tau)}{d\tau} =
 -1+w^2(\tau)\,.
\end{equation}
The solution of this equation with the initial condition $w(0)=1$ is of course
$w(\tau)=1$: the gas remains spatially homogeneous. Now we use Eq.~(\ref{pressure}) and obtain
\begin{equation}\label{puniform}
   \frac{\dot{p}(t)}{p(t)^{3/2}} = -1\,,
\end{equation}
which yields, in the rescaled variables, Haff's law for the gas pressure:
\begin{equation}\label{Haff}
    p(t)=\frac{1}{(1+t)^2}\,.
\end{equation}

The next test  of Eq.~(\ref{eq5}) is the linear stability analysis of a HCS. While the reduced Eq.~(\ref{eq5}) is not supposed to capture the evolution of small perturbations with an arbitrary polarization, it must reproduce correctly the evolution of the \textit{clustering} mode in the limit when the perturbation wavelengths are small compared with the sound travel distance $l_s$. Let us show
it to be indeed the case. We look for the solution of rescaled Eq.~(\ref{eq5}) in the form $w(m,\tau)=1+\delta
w(m,\tau)$, where $|\delta w(m,\tau)| \ll 1$. [Correspondingly,  the rescaled
density perturbation
$\delta \rho(m,\tau)=-2 \delta w (m,\tau).$] One can represent $\delta w(m,t)$ as a linear superposition of sines and
cosines of $k m$ with different (rescaled) wave numbers $k$. This fact, in conjunction
with the BCs at the ends of the channel, guarantees that
$\langle\delta w(m,\tau)\rangle=0$. Then Eq.~(\ref{eq5}) yields
\begin{equation}\label{growth1}
\frac{\partial}{\partial\tau} \delta w(m,\tau)=\delta
w(m,\tau)+\frac{\partial^2}{\partial m^2} \delta w(m,\tau)\,.
\end{equation}
For a single mode perturbation with wave number $k$ we obtain
\begin{equation}\label{growth2}
\delta w(m,\tau)=\delta w(m,0)\,e^{\hat{\Gamma}_k \tau}
\end{equation}
with the growth/damping rate
\begin{equation}\label{Gamma}
    \hat{\Gamma}_k=1-k^2\,.
\end{equation}
For $k<k_*=1$ (correspondingly, $k>k_*=1$)
Eqs.~(\ref{growth2}) and (\ref{Gamma}) describe an exponential growth
(correspondingly, decay) of a small single-mode perturbation in time $\tau$. Recalling that we rescaled the coordinate to the critical length $l_{cr}$, provided by the complete (unreduced) linear theory, we immediately notice that Eq.~(\ref{Gamma}) correctly predicts the instability threshold. To
go back to the physical time $t$ we substitute, in the leading order, Haff's
law~(\ref{Haff}) into Eq.~(\ref{tau}) and obtain, after elementary integration,
\begin{equation}\label{taut}
    \tau=\frac{1}{\gamma} \ln\left(1+t\right)\,.
\end{equation}
Plugging it into Eq.~(\ref{growth2}), we obtain an algebraic growth of the small
perturbations in the physical time:
\begin{equation}\label{growth3}
\delta w(m,t)=\delta w(m,0)\,\left(1+t\right)^{\hat{\Gamma}_k/\gamma} \,.
\end{equation}

\begin{figure}[ht]
\begin{tabular}{cc}
\includegraphics[scale=0.5]{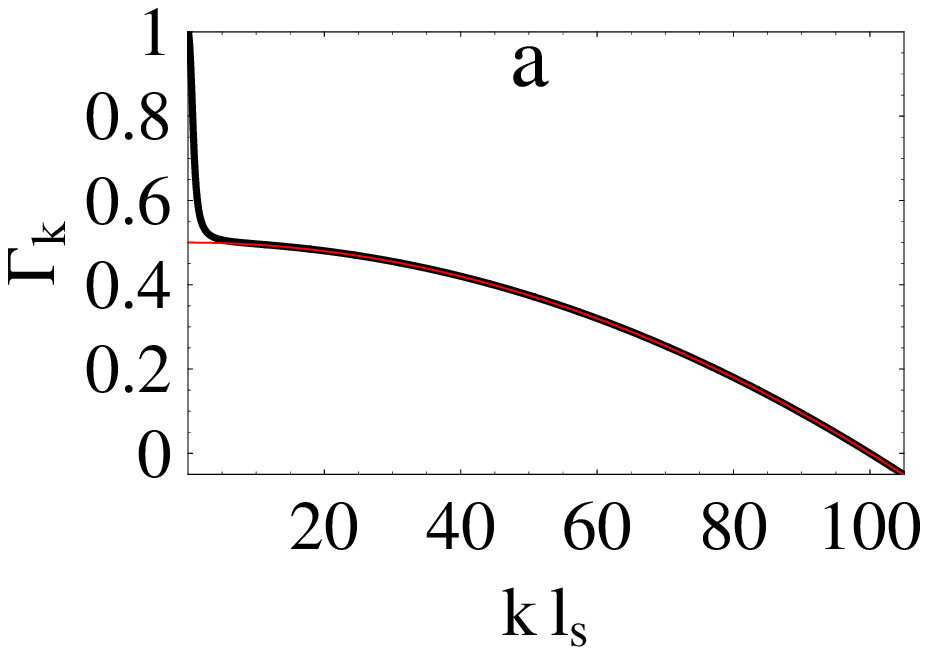}\\
\includegraphics[scale=0.5]{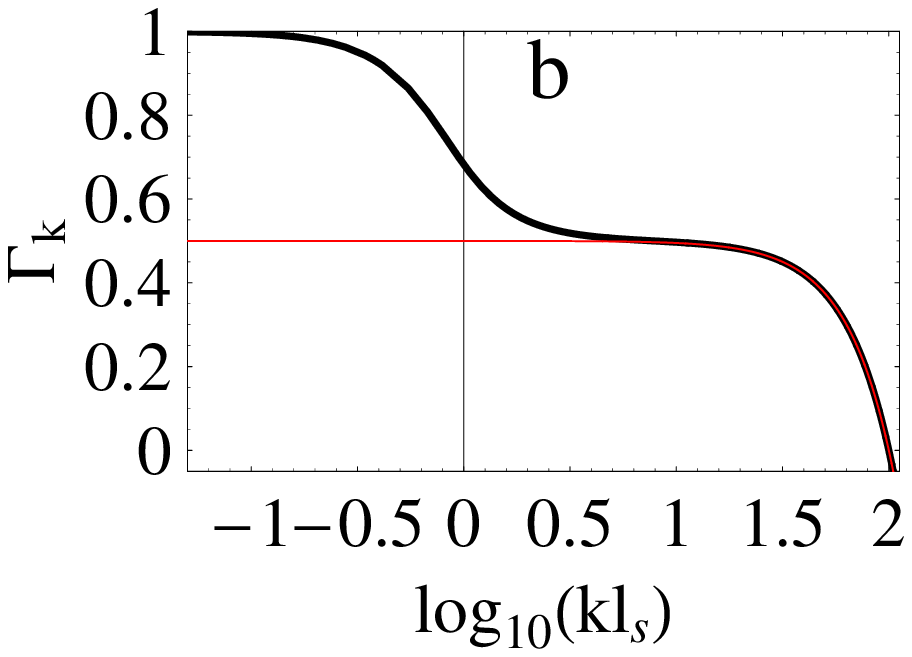}
\end{tabular}
\caption{(Color online.) The growth exponent $\Gamma_k$ of the clustering mode versus the rescaled wave number $k l_s$ (a) and versus $\log \,(k l_s)$ (b) as predicted from the complete linear stability analysis of a HCS \cite{McNamara1} (the thick black line) and from the reduced equation (\ref{eq5}) (the thin red line). The physical (not rescaled) units are used, and the parameters are $\gamma=2$ and $k_{cr}l_s= 100$. At $k
l_{s} \lesssim 1$  the isobaric growth rate (\ref{eq5}) underestimates the actual growth rate, but in the region of $k
l_{s} \gg 1$ excellent agreement between the two results is observed.} \label{lineargrowth}
\end{figure}

The growth exponent  $\Gamma=\hat{\Gamma}_k/\gamma$, with $\hat{\Gamma}_k$ from Eq.~(\ref{Gamma}) coincides with that obtained from the complete linear stability analysis \cite{McNamara1}, if we assume there $k
l_{s} \gg 1$ (in the physical units) and consider the clustering mode, rather than the two decaying
acoustic modes. Figure \ref{lineargrowth} shows this comparison in a graphic form. At $k
l_{s} \lesssim 1$  the isobaric growth rate underestimates the true growth rate, but in the region of $k
l_{s} \gg 1$ excellent agreement is observed. The comparison with the complete linear stability analysis is instructive for two more reasons. First, as was observed by McNamara \cite{McNamara1}, for $k l_{s} \gg 1$ the pressure perturbations of the clustering mode vanish in the leading order in $1/(k l_s)$. That is, the linear density and temperature perturbations grow on the background of an (almost) constant pressure. Second, the viscosity effects do not affect the growth exponent in this regime \cite{McNamara1}. As our reduced formalism shows,  the last two properties persist, for the low Mach number flow, in the nonlinear regime as well.

Having successfully tested our reduced model in these two simple cases, we now consider nonlinear evolution.

\section{Neglecting heat diffusion causes a density blowup}
\label{blowup}
As we mentioned earlier, the only governing parameter in Eq.~(\ref{eq5}), except parameters introduced by the initial condition, is the  rescaled system length/mass ${\cal L}$.  In the limit of
${\cal L}\gg 1$, and for a sufficiently large-scale initial condition, one can drop the diffusion term in Eq.~(\ref{eq5}). This approximation is valid as long as the solution remains large-scale.  At the level of linear stability analysis this (intermediate-wavelength) approximation is fully justified. Here Eq.~(\ref{Gamma}) becomes $\hat{\Gamma}_k\simeq 1$, and one is interested in the nonlinear development of the growing perturbations. With the diffusion term dropped we obtain
\begin{equation} \label{eq5inter}
 \frac{\partial w}{\partial \tau} =
 -1+ w\,\left\langle w\right\rangle\,.
\end{equation}
This nonlinear integro-differential evolution equation is exactly soluble for
any initial data $w(m,0)$. The complete solution is presented below.  The main result here is that, for any inhomogeneous initial condition, the solution
of Eq.~(\ref{eq5inter}) develops a zero $w$ (hence an infinite density) in a finite time.
Let us first discuss the properties of the solution in a close vicinity of the singularity $w\to 0$. In the leading order
we can neglect the integral term in Eq.~(\ref{eq5inter}) and obtain $\partial w/\partial \tau = -1$, so
that $w(m,\tau) = \tilde{w}(m)-\tau$, where $\tilde{w}(m)$ is a smooth function.
The singularity occurs in the Lagrangian point $m_0$ that corresponds to the
minimum of $\tilde{w}(m)$. The leading order behavior of the (rescaled) gas
density near the singularity is described by the following equation:
\begin{equation}\label{sing}
\rho(m,\tau) = \left[\tau_c-\tau+\frac{1}{2} \frac{d^2\tilde{w}}{d m^2}\left(m_0\right)
    (m-m_0)^2\right]^{-2}\,,
\end{equation}
where the time of singularity $\tau_c=\tilde{w}(m_0)$. The singularity structure, as described by Eq.~(\ref{sing}), coincides with that exhibited by a family of exact solutions of the full set of \textit{ideal} hydrodynamic equations [that is, Eqs.~(\ref{hydrodynamics1})-(\ref{hydrodynamics3}) without the viscous and heat diffusion terms], reported in Ref. \cite{Fouxon1,Fouxon2}. At $\tau=\tau_c$ the density blows up as
$\rho(m,\tau_c)\sim (m-m_0)^{-4}$. Going back to the Eulerian coordinate, we obtain a finite-mass density blowup
$\rho(x,\tau_c)\sim |x-x_0|^{-4/5}$, where $x_0$ is the Eulerian coordinate of the singularity.  We refer the reader to Ref. \cite{Fouxon2} for a detailed analysis of the structure of this singularity, as observed in the gas density, temperature and velocity. Notably, the pressure field does not have any singularity in the exact solutions \cite{Fouxon1,Fouxon2}, and is approximately constant in a narrow region around the density singularity. That is, the density blowup, as featured by the exact solutions of ideal granular hydrodynamics \cite{Fouxon1,Fouxon2},  locally obeys an isobaric scenario, as was noticed in Ref. \cite{Fouxon2}. This provides the reason why the same type of singularity appears in our reduced low Mach number theory.

Now we present a complete solution of Eq.~(\ref{eq5inter}). First, we obtain
a closed evolution equation for the (necessarily positive) quantity
$\chi(\tau)= \left\langle w(m,\tau) \right\rangle$
by integrating the both sides of Eq.~(\ref{eq5inter}) over $m$ from $0$ to ${\cal L}$:
\begin{equation}\label{Y1}
    \frac{d\chi(\tau)}{d\tau} = -1+\chi^2\,.
\end{equation}
We consider the solution of this equation with the initial condition
$$\chi_0 = \left\langle w(m,0) \right\rangle =
\left\langle\rho(m,0)^{-1/2}\, \right\rangle \le 1\,.
$$
The solution can be written as
\begin{equation}\label{Y2}
    \chi(\tau) = \frac{\chi_0 -
    \tanh (\tau)}{1-\chi_0 \tanh
    (\tau)}\,,
\end{equation}
Now we can rewrite Eq.~(\ref{eq5inter}) as
\begin{equation} \label{eq5a}
  \frac{\partial w}{\partial \tau} -\chi(\tau) \,w(m,\tau)=
 -1\,,
\end{equation}
where $\chi(\tau)$ is given by Eq.~(\ref{Y2}). Equation~(\ref{eq5a}) is easily soluble:
\begin{equation}\label{w}
    w(m,\tau)=\frac{w(m,0)+ \chi_0
    \left[\cosh\left(\tau\right)
    -1\right]
    -\sinh \left(\tau\right)}{\cosh \left(\tau\right) -
    \chi_0 \sinh \left(\tau\right)}\,.
\end{equation}
The presence of the factor $\chi_0 \left[\cosh\left(\tau\right)
-1\right] -\sinh \left(\tau\right)$ in the numerator of Eq.~(\ref{w}) causes,
for any (non-constant) initial data $w(m,0)$, a singularity $w\to 0$ in a finite
time. The singularity occurs at the Lagrangian point $m_0$ where the function
$w(m,0)$ has its minimum, at time
\begin{equation}\label{tau*}
    \tau_c = \ln \left[\frac{\Delta w+\left(\Delta w^2+1-\chi_0^2\right)^{1/2}}
    {1-\chi_0}\right]\,
\end{equation}
where $\Delta w \equiv w(m_0,0)-\chi_0$. Note that $\Delta w=(1/{\cal L})\int_0^{{\cal L}} [w(m_0, 0)-w(m,\tau)]
dm\,\leq 0$.

Now we compute the (rescaled) pressure $p(\tau)$ from Eq.~(\ref{pressure1}) [note that the
right hand side is simply $\chi(\tau)$ given by Eq.~(\ref{Y2})],
\begin{equation}\label{pressure3}
    p(\tau)=\left(\cosh\tau -
  \chi_0 \sinh\tau\right)^{2\gamma}\,,
\end{equation}
and use this result in Eq.~(\ref{t}) for the rescaled physical time:
\begin{equation}\label{t1}
    t=\gamma \int_0^{\tau}
    \frac{d\tau^{\prime}}{\left(\cosh \tau^{\prime} - \chi_0 \sinh
    \tau^{\prime}\right)^{\gamma}}\,.
\end{equation}
For $\gamma=2$ (a 2D gas of disks) this integral is
elementary, and the result is
\begin{equation}\label{t2}
   t= \frac{2}{\coth\tau-\chi_0}\,.
\end{equation}
Now we can express $\tau$ through $t$,
\begin{equation}\label{tau1}
\tau = \mbox{arccoth} \left(\frac{2}{t}+\chi_0\right)\,
\end{equation}
and rewrite Eqs.~(\ref{w}) and (\ref{pressure3}) (for $\gamma=2$) as
\begin{eqnarray}
 w(m,t)&=& \frac{1}{2} \{\left[w(m,0)-\chi_0\right]\sqrt{4+4
\chi_0 t -
  (1-\chi_0^2)t} \nonumber \\
&+& 2 \chi_0 - (1-\chi_0^2) t\}\,. \label{w1}
\end{eqnarray}
and
\begin{equation}\label{pressure4}
p=\left[1+\chi_0 t -
\left(1-\chi_0^2\right)\frac{t^2}{4} \right]^{-2}\,.
\end{equation}
So, the solution for $\gamma=2$ is surprisingly simple. We remind
that, in view of the chosen rescaling, the initial condition $w(m,0)$
must obey $\left \langle w^2(m,0)\right\rangle=1$. To return to the HCS and Haff's law in Eqs.~(\ref{w1}) and (\ref{pressure4}) one should put there $w(m,0)=\chi_0=1$. Equation~(\ref{pressure4}) shows that Haff's law is an upper bound for
the thermal energy loss rate: any deviation from homogeneity
brings about $\chi_0<1$  and a slower thermal energy decay.

Let us note that the solution (\ref{Y2}) for $\chi(\tau)$ vanishes at
$\tau_*=(1/2)\,\ln [(1+\chi_0)/(1-\chi_0)]$ and becomes negative at larger $\tau$. This is in apparent
contradiction with the positivity of $w$ that dictates $\chi(\tau)=\left\langle w(m,\tau) \right\rangle\geq 0$.
The contradiction is resolved by noting that $\tau_*$ is always greater than the singularity time $\tau_c$, beyond
which the solution does not apply. [To see that $\tau_c\leq \tau_*$ one can use, in Eq.~(\ref{tau*}),  that $\Delta w+\left(\Delta w^2+1-\chi_0^2\right)^{1/2}
\leq \left(1-\chi_0^2\right)^{1/2}$ for any $\Delta w\leq 0$.]
Similarly, the pressure as predicted by Eq.~(\ref{pressure3})
or Eq.~(\ref{pressure4}) would start increasing at some time. At physically meaningful times
$\tau<\tau_c$, however, we have $\chi(\tau)>0$, and the pressure always decreases in accord with Eq.~(\ref{pressure1}).

\begin{figure}[t]
\begin{tabular}{ccc}
\includegraphics[scale=0.5]{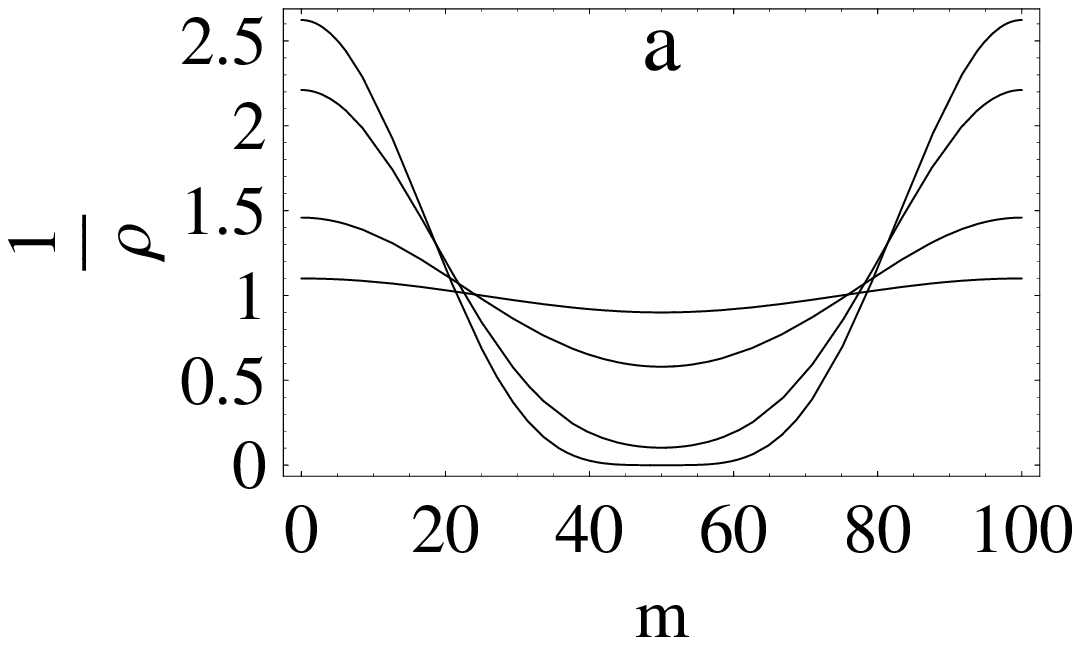}\\
\hspace{0.3cm}\includegraphics[scale=0.51]{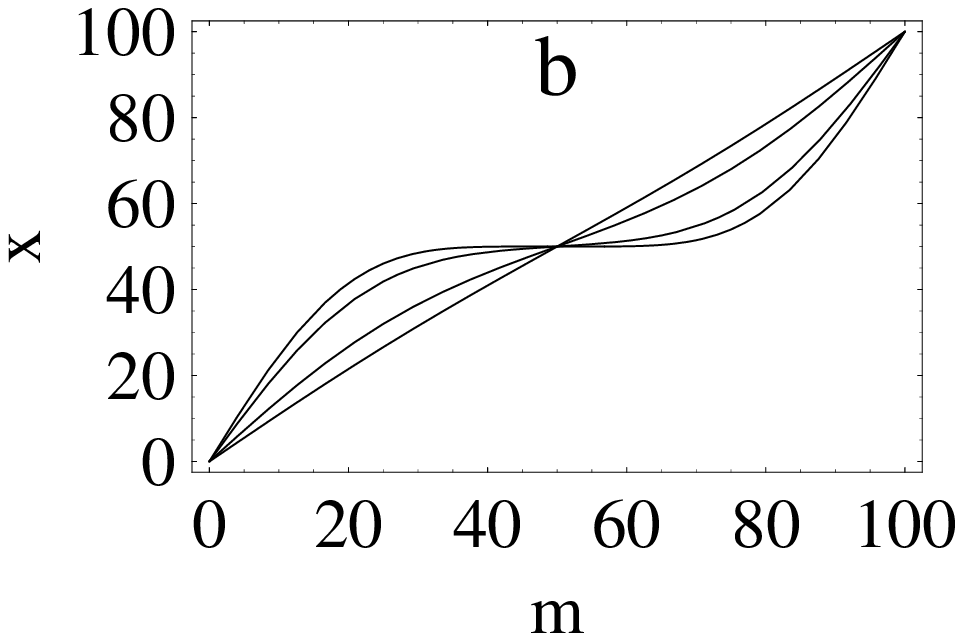}\\
\includegraphics[scale=0.5]{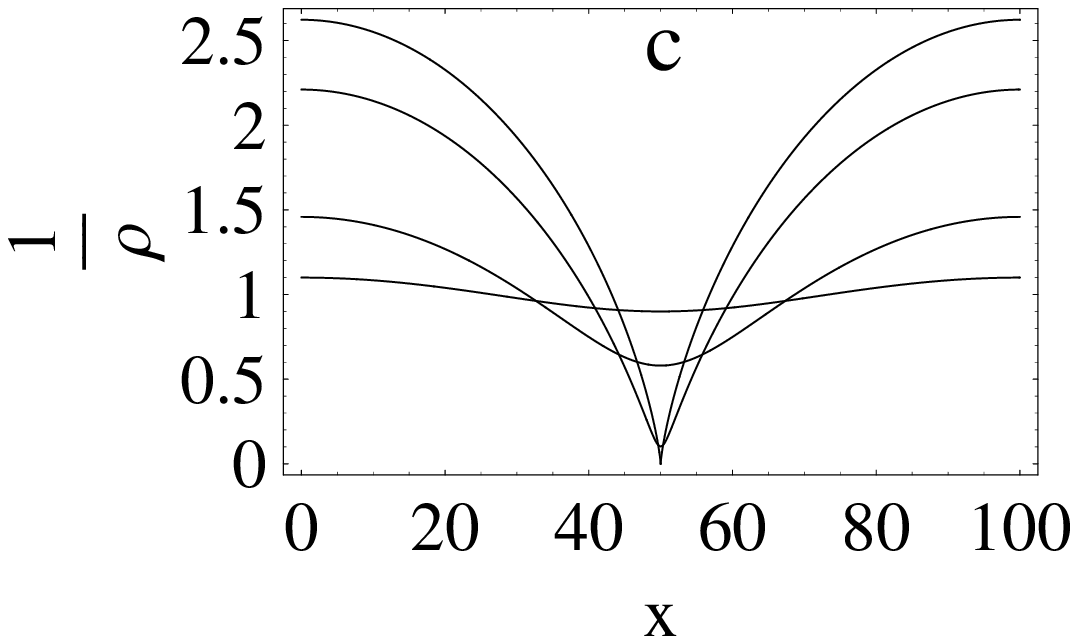}
\end{tabular}
\caption{The density history of a freely cooling gas of inelastic hard disks in a 2D channel in the zero-heat-diffusion limit. The rescaled initial density $\rho(m,0)=\left[1+0.1 \cos (2\pi m/{\cal L})\right]^{-1}$. The rescaled system length/mass
${\cal L} =100$. Panel a: the rescaled inverse density of the gas, $1/\rho$, versus the
Lagrangian mass coordinate $m$ at times $\tau=0$, $1.5$, $2.5$ and the time of
singularity $\tau_c \simeq 2.8755$. Panel b: the rescaled Eulerian
coordinate $x$ versus $m$ at the same
times. Panel c: $1/\rho$ versus $x$ at the same times. The sequence of curves is self-explanatory.} \label{um_xm_ux}
\end{figure}

As a simple illustration of our solution (\ref{w}), let us chose the following initial condition: $w(m,0)=\left[1+\delta \cos
(2 \pi m/{\cal L})\right]^{1/2},\, 0<\delta<1$. In this case
\begin{eqnarray}
 \nonumber
\chi &=& \frac{1}{\pi} \left[\sqrt{1-\delta }\; \mbox{E}\left(\frac{2 \delta
}{\delta
   -1}\right)+\sqrt{\delta +1} \;\mbox{E}\left(\frac{2 \delta }{\delta
   +1}\right)\right]\,,
\end{eqnarray}
where $\mbox{E}(\dots)$ is the complete elliptic integral of the second kind, see \textit{e.g.} \cite{Abramowitz}. Figure~\ref{um_xm_ux}a shows, at different times, the rescaled inverse density
$1/\rho(m,\tau)$, as obtained from Eq.~(\ref{w}), for $\delta=0.1$ and  ${\cal L}=100$. Figure~\ref{um_xm_ux}b depicts, at the same times, the rescaled Eulerian coordinate
$x = \int_0^m w^2(m^{\prime},\tau) dm^{\prime}$ versus the Lagrangian coordinate $m$.
Figure~\ref{um_xm_ux}c shows the rescaled inverse density in the rescaled
Eulerian coordinates and illustrates the emergence of the cusp density singularity at $x={\cal L}/2$. The inverse density behaves like $(m-m_0)^4$ at small $m-m_0$ in the Lagrangian coordinate, and like $(x-x_0)^{4/5}$ at small $x-x_0$ in the Eulerian coordinate. This simple example is instructive as, for $\delta \ll 1$, this initial condition
corresponds to a small single-mode density perturbation, so the initial
evolution is describable by the linear theory.

\section{Heat diffusion arrests the density blowup}
\label{arrest} A central result of this work is in that, no matter how small
initially, the heat diffusion term in Eq.~(\ref{eq5}) arrests the density
blowup.  An emerging balance of the inelastic cooling and heat diffusion leads
to existence of steady state solutions of Eq.~(\ref{eq5}). These solutions
describe novel cooling states of the granular gas, where the (inhomogeneous)
density profile is time-independent, while the (homogeneous) pressure continues
to decay with time. We found that, in our rescaled variables, the density
profile of the novel cooling state is uniquely defined by the parameter ${\cal
L}$.  For sufficiently large values of the rescaled length/mass, ${\cal L} \gg
1$, the maximum gas density of the novel cooling state is exponentially large in
${\cal L}$. In the low Mach number theory, considered in this work, the novel
cooling states represent global attractors, as they develop for any
inhomogeneous initial conditions. Finally, the novel cooling states represent
\textit{exact} solutions of the complete, unreduced set of hydrodynamic
equations (\ref{hydrodynamics1})-(\ref{hydrodynamics3}).

\subsection{Steady state density profiles}
Steady-state solutions of Eq.~(\ref{eq5}) are described by the
equation
\begin{equation} \label{eq51}
\frac{d^2 w}{d m^2}=w-\left\langle w\right\rangle w^2\,.
\end{equation}
Notice that, although obtained from our reduced, low Mach number theory,
Eq.~(\ref{eq51}) also follows from the full set of hydrodynamic equations (\ref{hydrodynamics1})-(\ref{hydrodynamics3}), if one assumes a homogeneous pressure and zero fluid velocity, and transforms to the Lagrangian coordinates.

Equation~(\ref{eq51}) is defined on the interval $0\le m\le{\cal L}$, at the ends of which we demand either periodic, or no-flux (zero first derivative) BCs. The solutions we are interested in must obey the conservation law (\ref{conslaw}). To get rid of the (\textit{a priori} unknown) factor $\langle w \rangle$, we introduce a new variable
\begin{eqnarray}&&
f(m)=\left\langle w\right\rangle w(m)
\end{eqnarray}
and obtain\begin{equation}\label{ordinary}
   \frac{d^2 f}{dm^2}-f+f^2=0\,.
\end{equation}
Once $f$ is found, one can restore $w$ via
\begin{equation}
\label{old} w=\frac{f}{\sqrt{\left\langle f\right\rangle}}.
\end{equation}
The conservation law (\ref{conslaw}) enforces a normalization condition
\begin{eqnarray}&&
\langle f^2\rangle=\langle f\rangle\,
\end{eqnarray}
that, in virtue of Eq.~(\ref{ordinary}), is obeyed automatically for the periodic or no-flux BCs.

Equation~(\ref{ordinary}) has appeared in numerous applications, and its solutions are well known.  It is convenient to interpret $f$ as a coordinate of a
Newtonian particle of unit mass, moving in a potential $U(f)=f^3/3-f^2/2$. The
``total energy" $E$ is conserved:
\begin{eqnarray}&&
E=\frac{1}{2}\left(\frac{df}{dm}\right)^2+\frac{f^3}{3}-\frac{f^2}{2}.\label{energyintegral}
\end{eqnarray}
For the bounded (spatially oscillating) solutions, $-1/6\leq E\leq 0$, we can write
\begin{equation}
\frac{f^3}{3}-\frac{f^2}{2}-E=\frac{(f-a)(f-b)(f-c)}{3},
\end{equation}
where $a>b>c$ are the real roots of the cubic polynomial.
Then the bounded solutions of Eq.~(\ref{ordinary}) can be written as
\begin{eqnarray}&&
f(m)=c+(a-c)\,{\mbox dn}^2\left(\sqrt{\frac{a-c}{6}}m,\, s\right)\,,
\label{solution}
\end{eqnarray}
where
\begin{equation}\label{s}
    s=\frac{a-b}{a-c}\,,
\end{equation}
and ${\mbox dn}$ is one of the Jacobi elliptic functions, see \textit{e.g.} \cite{Abramowitz}.
There are two limits when Eq.~(\ref{solution}) simplifies. In the limit of $E=-1/6+\delta E$, $0<\delta E \ll 1$,
the solution, $f(m)=1+\sqrt{2\delta E} \cos m$, corresponds to a small-amplitude sinusoidal modulation of the HCS $w(m)=1$. In the limit of $E\to 0$, we
have $a=3/2$ and $b=c=0$, so that
\begin{equation}
\label{cosh} f(m, E\to 0)=\frac{3}{2}\,{\mbox dn}^2\left(\frac{m}{2},
1\right)=\frac{3}{2}\, \cosh^{-2}\left(\frac{m}{2}\right),
\end{equation}

Using Eqs.~ (\ref{old}) and (\ref{solution}), we rewrite the steady state
solutions in terms of $w(m)$:
\begin{equation}\label{w(m)}
    w(m)=\frac{c+(a-c)\, {\mbox dn}^2\left(\sqrt{\frac{a-c}{6}}\,m, s\right)}
    {\sqrt{c+(a-c)\,\frac{\mbox{E}(s)}{\mbox{K}(s)}}}\,,
\end{equation}
where $\mbox{K}(s)$ is the complete elliptic integral of the first kind. The
lagrangian spatial period, or wavelength, of the solution (\ref{w(m)}) is
\begin{equation}\label{wavelength}
    \Pi=\sqrt{\frac{24}{a-c}}\, \mbox{K}(s)\,.
\end{equation}
In the limit of $s\to 0$ (or $E\to -1/6$), the wavelength (\ref{wavelength}) reaches its minimum
value $2\pi$. If the rescaled channel length
${\cal L}$ is less than $2\pi$ (for the periodic BCs), or less than
$\pi$ (for the no-flux BCs), the only possible steady state is the constant
density state $w(m)=1$ corresponding to Haff's law. This result is in full agreement with the linear stability analysis of Eq.~(\ref{eq5}), see Eq.~(\ref{Gamma}). When
${\cal L}$ exceeds $2 \pi$ (for the periodic BCs), or $\pi$ (for the no-flux BCs), the HCS bifurcates into an inhomogeneous steady state (\ref{w(m)}). In general, the rescaled channel length/mass ${\cal L}$ must be equal, by virtue of the BCs,
to an integer number of $\Pi$ (for the periodic BCs), or to an
integer number of $\Pi/2$ (for the no-flux BCs). For
sufficiently large value of ${\cal L}$, therefore, a whole \textit{family} of steady state density
profiles exists. Which of the steady state solutions is selected by the dynamics of
Eq.~(\ref{eq5})?

\subsection{Selected steady-state solutions: the inhomogeneous cooling states}

\begin{figure}[ht]
\includegraphics[scale=0.85]{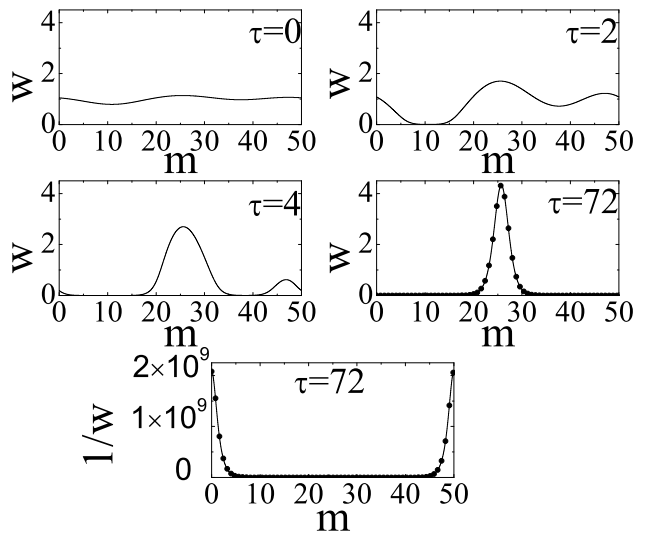}
\caption{Numerical $w$-profiles at times $\tau=0$, $2$, $4$ and $72$,  and $1/w$-profile at time $\tau=72$, for ${\cal L}=50$ when starting from the initial condition (\ref{incond}).  The two panels for $\tau=72$ also show, by circles, the single hole asymptotes (\ref{cosh1}) and (\ref{cosh2}), respectively.} \label{snapshots1}
\end{figure}

We performed extensive numerical simulations with Eq.~(\ref{eq5}), using a specially developed numerical scheme described in Appendix A. Both periodic, and no-flux BCs were used. We observed that, when $0<{\cal L}<2\pi$ (for the periodic BCs), or $0<{\cal L}<\pi$ (for the no-flux conditions), the HCS appears, as expected.  When ${\cal L}$ exceeds $2\pi$ (for the periodic BCs), a weakly inhomogeneous steady state density profile sets in. As ${\cal L}$ increases further, the weakly inhomogeneous states develops into a strongly inhomogeneous states.
The simulations showed that the rescaled length/mass of the gas, ${\cal L}$, uniquely selects the emerging steady state density profile, while the initial $w$-profile does not play any role in the selection. For a given ${\cal L}$ the dynamics always selects, out of the family of
steady state solutions (\ref{w(m)}), the one with the \textit{maximum} possible wavelength $\Pi$:
\begin{equation}\label{maxwavelength}
{\cal L}=\left\{\begin{array}{ll}
\Pi
& \mbox{for the periodic BCs}\,, \\
\Pi/2 & \mbox{for the no-flux BCs\,.}
\end{array}
\right.
\end{equation}
Snapshots from a typical simulation (one of many that we performed) for the periodic BCs are shown in Fig. \ref{snapshots1}. The initial condition is this example was
\begin{eqnarray}
\nonumber  w^2(m,0)&=& 1-0.1\,\cos (2\pi m/{\cal L})- 0.15\, \sin (2\pi m/{\cal L})  \\
  \label{incond}
  &+& 0.2 \, \cos(4\pi m/{\cal L}) -0.05 \sin (4\pi m/{\cal L})\,.
\end{eqnarray}
The rescaled system length/mass ${\cal L}=50$ was sufficiently large to fit in steady state solutions with several oscillations. Nevertheless, the dynamics selected the solution with the spatial period equal to the rescaled system length ${\cal L}$.

\begin{figure}[ht]
\begin{tabular}{cc}
\includegraphics[scale=0.7]{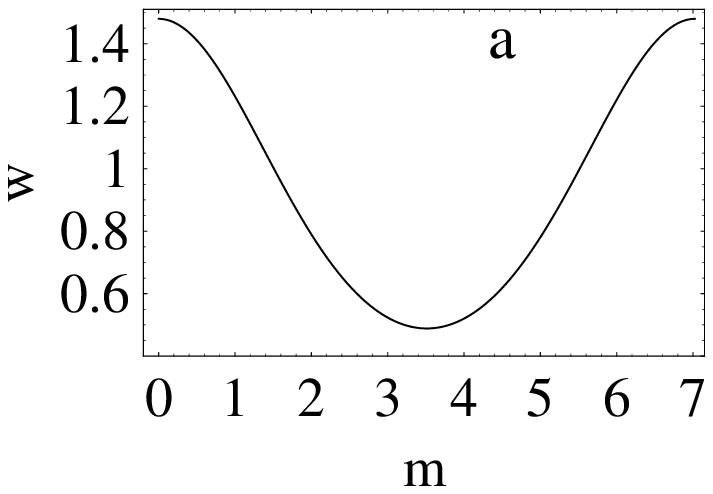}\\
\includegraphics[scale=0.7]{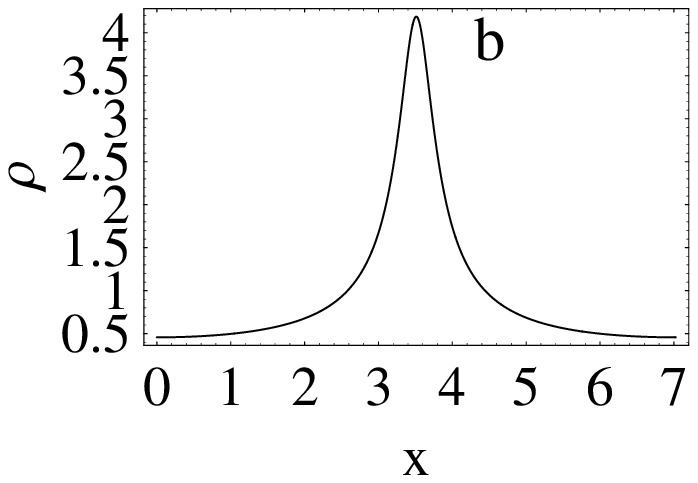}\\
\end{tabular}
\caption{The inhomogeneous cooling state for ${\cal L}=7.025$. Panel a: the Lagrangian steady state solution $w(m)$ as predicted by Eq.~(\ref{w(m)}). Panel b: the rescaled steady state gas density $\rho$ versus the rescaled
Eulerian coordinate $x$.} \label{eps0.8}
\end{figure}

\begin{figure}[ht]
\begin{tabular}{cc}
\includegraphics[scale=0.7]{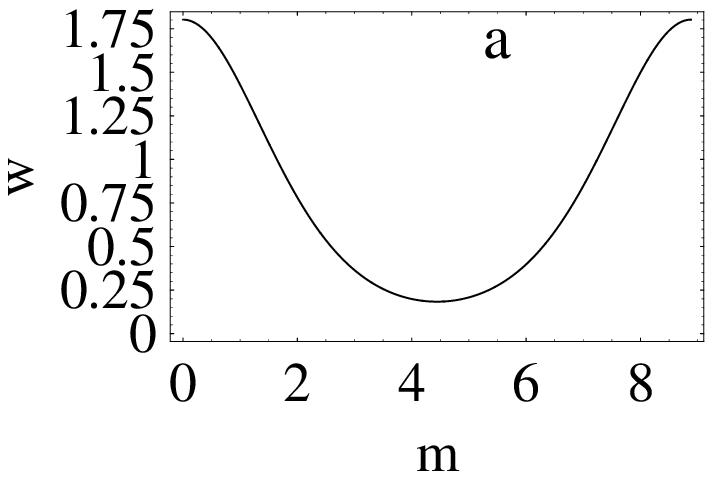}\\
\hspace{0.2cm}
\includegraphics[scale=0.72]{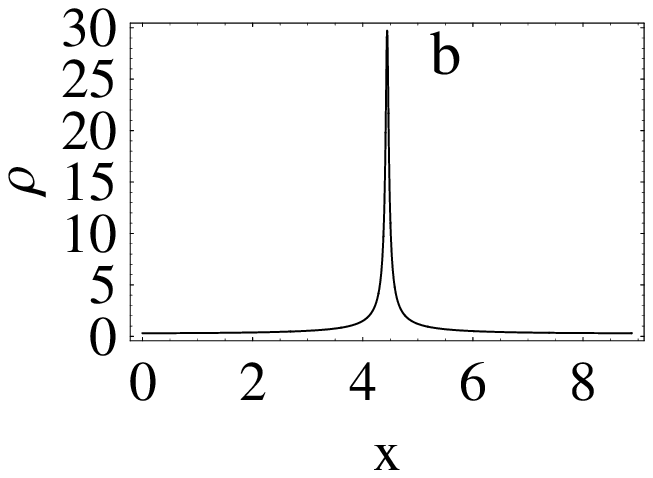}\\
\end{tabular}
\caption{Same as in Fig.~\ref{eps0.8}, but for ${\cal L}=8.886$.} \label{eps0.5}
\end{figure}

\begin{figure}[ht]
\begin{tabular}{cc}
\includegraphics[scale=0.7]{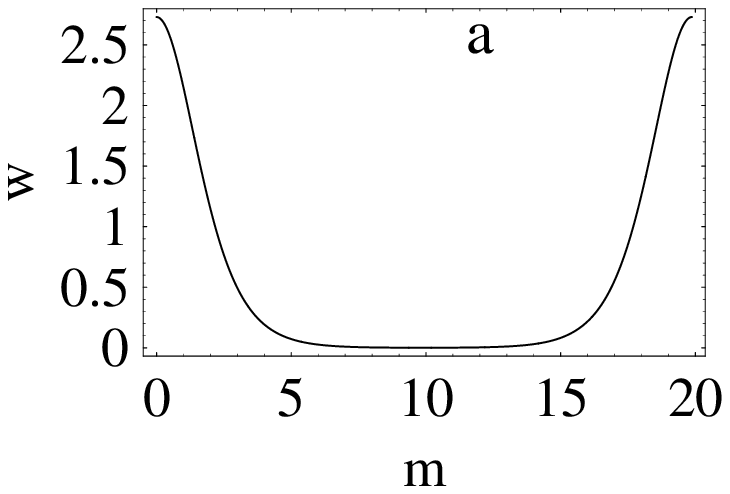}\\
\hspace{-0.4cm}
\includegraphics[scale=0.75]{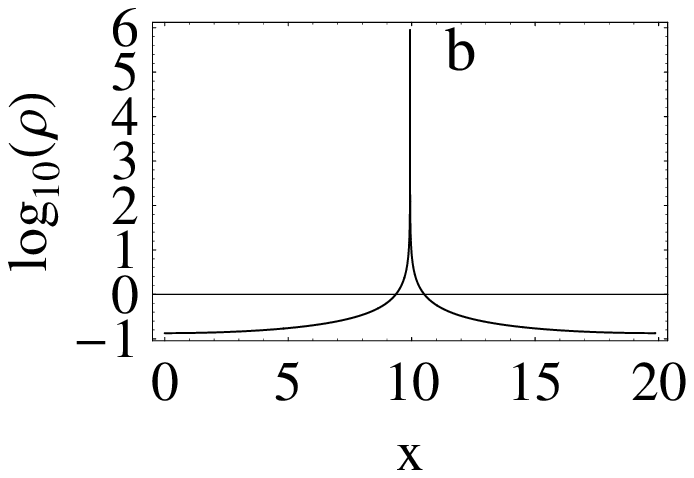}\\
\end{tabular}
\caption{Same as in Fig.~\ref{eps0.8}, but for ${\cal L}=19.869$. Notice the logarithmic scale in panel b.} \label{eps0.1}
\end{figure}

Figures \ref{eps0.8} - \ref{eps0.1} depict our analytical solutions (\ref{w(m)}) in the Lagrangian coordinate, and the corresponding density profiles in the
Eulerian coordinates, for three different values of the parameter
${\cal L}$. Here we assumed the periodic BCs and
(arbitrarily) chose the position of the minimum of $w(m)$ to be in the middle of
the channel.

\begin{figure}[ht]
\begin{tabular}{cc}
\includegraphics[scale=0.7]{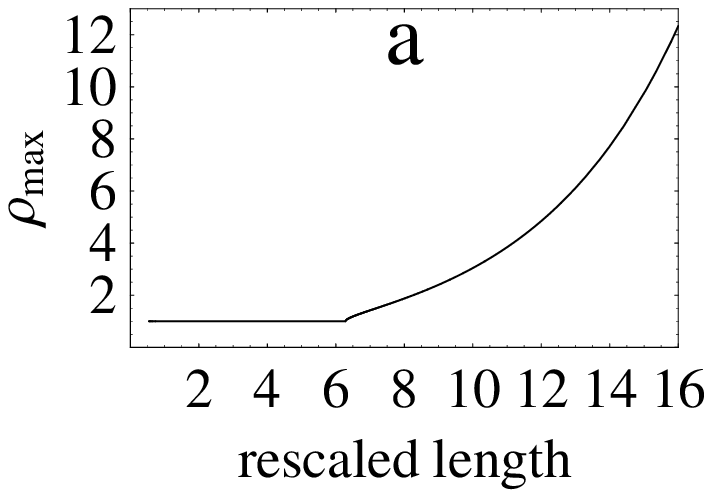}\\
\hspace{-0.4cm}
\includegraphics[scale=0.75]{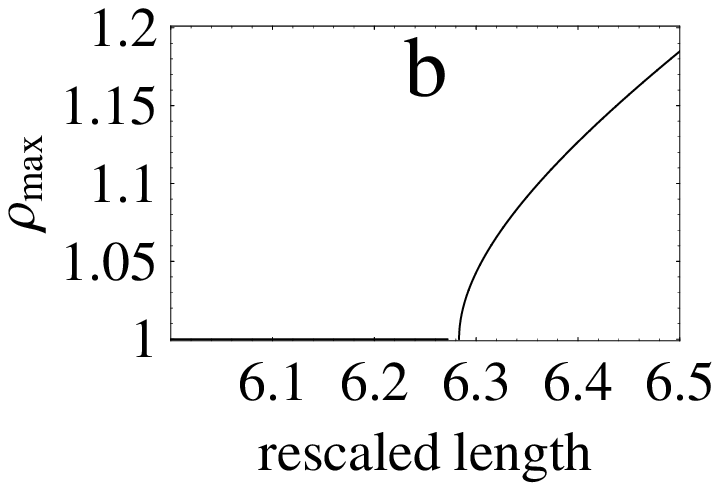}\\
\end{tabular}
\caption{The bifurcation diagram of the freely cooling granular gas in a channel. Shown is the maximum (rescaled) steady state density of the gas versus the rescaled channel length ${\cal L}$, predicted by Eqs.~(\ref{w(m)}) and (\ref{wavelength}). Panel b focuses on a vicinity of the  supercritical bifurcation point  ${\cal L} =2\pi$.} \label{maxdens}
\end{figure}

The maximum (rescaled) gas density versus the rescaled channel length ${\cal L}$, predicted by Eqs.~(\ref{w(m)}) and (\ref{wavelength}), is shown in Fig.~\ref{maxdens}. This dependence can serve
as a bifurcation diagram of the system. One observes, at ${\cal L} > 2 \pi$, a supercritical bifurcation from the HCS to an ICS.

One can see that, as the parameter ${\cal L}$
increases, the maximum gas density in the cluster grows very fast [note that Fig.~\ref{eps0.1}b shows the density in logarithmic scale]. Let us consider the asymptotic form of the solution at ${\cal L}\gg 1$ in some detail.
The density
maximum in this case is \textit{exponentially} large \cite{nottoolong}. This is due to the behavior of the $s\to 1$
asymptotics of the steady-state solution, see Eq.~(\ref{cosh}). In this case the ``energy" $E$ is very small, and can be expressed through the rescaled system length as $|E|\simeq 72\,\exp(-{\cal L})$. The maximum value of $w(m)$ is
\begin{equation}\label{wmax}
    w_{max}\simeq\sqrt{\frac{3{\cal L}}{8}}\,.
\end{equation}
To obtain the minimum value of $w(m)$ (that corresponds to the maximum value of the density), it is convenient to use the exact relation $w_{min}=b/\sqrt{\langle f \rangle}$ and calculate the asymptotic value of $b$ at $|E|\ll 1$, or ${\cal L}\gg 1$. The result is
\begin{equation}\label{wmin}
    w_{min}\simeq\sqrt{24 {\cal L}}\,
    e^{-{\cal L}/2}\,,
\end{equation}
By virtue of Eq.~(\ref{cosh}), the asymptotics of the steady state solution (\ref{w(m)}) at $|m|\ll {\cal L}/2$ is
\begin{equation}\label{cosh1}
    w_0(m)\simeq\sqrt{\frac{3{\cal L}}{8}}\,\cosh^{-2}
    \left(\frac{m}{2}\right)\,,
\end{equation}
where, for convenience, we have written the solution on the interval $-{\cal L}/2<m<{\cal L}/2$ and used the approximate equality $\langle f \rangle \simeq 6/{\cal L}$. To calculate the asymptotics of Eq.~(\ref{w(m)}) at $|m| \gg 1$, we can deal directly with Eq.~(\ref{ordinary}) and neglect the $f^2$ term. The solution of the resulting elementary equation is a linear combination of $e^{m}$ and $e^{-m}$. The two arbitrary constants can be determined from the two conditions at $|m|={\cal L}/2$:  $df/dm=0$ and $w_0\equiv f/\sqrt{\langle f \rangle}=w_{min}$, where $w_{min}$ is given by Eq.~(\ref{wmin}). We obtain
\begin{equation}\label{cosh2}
    w_0(m)\simeq\sqrt{24{\cal L}}\,e^{-{\cal L}/2}\cosh ({\cal L}/2-|m|)\,,\;\;\; |m|\gg 1\,.
\end{equation}
Note that the asymptotes (\ref{cosh1}) and (\ref{cosh2}) coincide in their common region $1\ll |m|\ll {\cal L}/2$, where each of them yields
\begin{equation}\label{expon}
    w_0(m)\simeq \sqrt{6{\cal L}}\,
   e^{-m}\,.
\end{equation}
Note that $\langle w_0(m) \rangle \simeq \sqrt{6/{\cal L}}$ is determined by the asymptote (\ref{cosh1}). We compared the asymptotes (\ref{cosh1}) and (\ref{cosh2}) with the numerical solution, shown in Fig.~\ref{snapshots1}, at a late time $\tau=72$. Employing the periodic BCs, we shifted the numerical solution in $m$ so that the maximum of $w(m, \tau=72)$ is at $m=0$. One can see that the agreement is excellent.

As higher $w$ corresponds to a lower gas density, the region of the maximum of
$w$ corresponds to a \textit{hole} in the density.
Therefore, we will call the approximate solution, fully determined by Eqs.~(\ref{cosh1}) and (\ref{cosh2}), the hole solution.
The rescaled steady-state gas
density, in the limit of ${\cal L} \gg 1$, is
\begin{equation}\label{denscosh1}
   \rho(m)\simeq\frac{8}{3 {\cal L}}\,\cosh^{4}
    \left(\frac{m}{2}\right)\,,\;\;\;|m|\ll {\cal L}/2\,,
\end{equation}
and
\begin{equation}\label{denscosh2}
\rho(m)\simeq\frac{e^{\cal L}}{24 {\cal L}}\,\cosh^{-2}
\left(\frac{{\cal L}}{2}-|m|\right)\,,\;\;\;|m| \gg 1\,,
\end{equation}
and the maximum and minimum density values are
\begin{equation}\label{maxmindens}
   \rho_{max}\simeq\frac{e^{{\cal L}}}{24 {\cal L}}\,,\;\;\;\;\;\;
   \rho_{min}\simeq\frac{8}{3 {\cal L}}\,.
\end{equation}
Note that Eqs.~(\ref{wmax})-(\ref{maxmindens}) work very well
already for moderate values of ${\cal L}$. For the dilute hydrodynamics to be still valid in the gas density peak
region, we must demand that the peak density be much less than the close packing
density. In view of the exponential growth of the maximum density with the
parameter ${\cal L}$, see Eq.~(\ref{maxmindens}), this leads to a
stringent condition:
$$
\rho_0\sigma^d \ll 24 {\cal L}\, e^{-{\cal L}} \,.
$$
If this condition is not fulfilled, the dilute theory will break down, and the attempted density blowup will be regularized by close-packing effects.

The general form of the steady state density profile in the Eulerian coordinates is quite cumbersome. However, its asymptotic form at ${\cal L} \gg 1$ that corresponds to the Lagrangian profiles (\ref{cosh1}) and  (\ref{cosh2}) is both elementary and instructive. For  Eq.~(\ref{cosh1}) one finds, after some algebra,
\begin{equation}\label{cosheuler}
    w_0(x)=\sqrt{\frac{3 {\cal L}}{2}}\,\cos\left[\frac{1}{3} \arccos \left(1-\frac{8 x^2}{{\cal L}^2} \right)\right] - \sqrt{\frac{3 {\cal L}}{8}}\,.
\end{equation}
This asymptotics is valid at $e^{-{\cal L}} \ll 1-2|x|/{\cal L}$, that is
almost over the whole channel $|x|<{\cal L}/2$ except in
a narrow region.  This region, however, includes a significant part of the gas mass, as evidenced by the size of this region in the Lagrangian coordinate and by the non-integrable diverging power-law asymptotics of the gas density:
\begin{equation}\label{common}
    \rho(x)
    \simeq \frac{1}{{\cal L}-2|x|}\;\;\;
\mbox{at}\;\;\;e^{-{\cal L}} \ll 1-\frac{2 |x|}{{\cal L}}\ll 1\,.
\end{equation}
There is of course no actual density divergence here, as Eq.~(\ref{common}) does not hold close to the end points:  at $1-2|x|/{\cal L} \lesssim e^{-{\cal L}}$. To find the density profile in this exponentially narrow region, we
express the relation between $x$ and $m$ as
\begin{eqnarray}
\nonumber  x &=& \int_0^m w_0^2(m^{\prime}) \,dm^{\prime}  \\
\nonumber &=& \int_0^{{\cal L}/2} w_0^2(m^{\prime}) \,dm^{\prime}  - \int_m^{{\cal L}/2} w_0^2(m^{\prime}) \,dm^{\prime} \\
\label{euler10}&=& {\cal L}/2 - \int_m^{{\cal L}/2} w_0^2(m^{\prime}) \,dm^{\prime}\,.
\end{eqnarray}
This form is convenient in the vicinity of $m={\cal L}/2$. The case of $m=-{\cal L}/2$ can be treated similarly, and the expressions that follow are valid in both cases. For $|m|\gg 1$  Eqs.~(\ref{euler10}) and (\ref{cosh2}) yield
\begin{equation}\label{euler11}
    |x|\simeq\frac{{\cal L}}{2}-6\,{\cal L} e^{-{\cal L}} \left[{\cal L}-2 |m|+\sinh({\cal L}-2 |m|) \right]\,.
\end{equation}
Equations~(\ref{denscosh2}) and (\ref{euler11}) determine, in a parametric form and in elementary functions, the density profile in the region sufficiently  far from the density minimum. Still simpler results can be obtained in the following two sub-regions. The first is the common region  ${\cal L}/2-|m|\gg 1$  but $|m|\gg 1$. The  asymptotics of Eqs.~(\ref{denscosh2}) and (\ref{euler11}) at ${\cal L}/2-|m| \gg 1$ become $\rho=e^{2|m|}/(6 {\cal L})$, and $|x|\simeq {\cal L}/2-3 {\cal L} e^{-2|m|}$, therefore $\rho=({\cal L}-2 |x|)^{-1}$ which coincides with
the asymptotics (\ref{common}) of Eq.~(\ref{cosheuler}).  The second limit corresponds to ${\cal L}/2-|m| \ll 1$. Here Eq.~(\ref{denscosh2}) becomes
$$
\rho(m)\simeq \frac{e^{{\cal L}}}{24 {\cal L}}\left[1-\left(\frac{{\cal L}}{2}-|m|\right)^2\right] \,,
$$
whereas Eq.~(\ref{euler11}) yields $|x|={\cal L}/2-24 {\cal L} e^{-{\cal L}} ({\cal L}/2-|m|)$. The resulting Eulerian density profile is
\begin{equation}\label{euler12}
\rho(x)\simeq \frac{e^{{\cal L}}}{24 {\cal L}}\left[1-\left(\frac{e^{{\cal L}}}{24 {\cal L}}\right)^2 \left(\frac{{\cal L}}{2}-|x|\right)^2\right]\,.
\end{equation}
\subsection{Energy decay for the ICSs}

Now let us consider the evolution of the (rescaled) total energy of the gas,
\begin{equation}
E_{tot}(t)=\int_{0}^{{\cal L}}  \left(\frac{p}{\gamma-1}+\frac{\rho v^2}{2} \right)dx\,,
\end{equation}
where the first term under the integral is the thermal energy density, and the
second term is the macroscopic kinetic energy density. For the low Mach number
flow, that we are dealing with in this work, the first term is almost independent of $x$, while the
second term is negligible. As a result, the energy decays, in the leading order, in the same way as the pressure. The pressure decay $p(\tau)$ is described by  Eq.~(\ref{pressure1}), whereas to go back to the physical time we use Eq.~(\ref{t}). For our steady state solutions we arrive at a generalized Haff's law
\begin{equation}
p(t)=\frac{1}{(1+\langle w\rangle t)^2}. \label{Haffgen}
\end{equation}
As $\langle w\rangle \leq \langle w^2\rangle^{1/2}=1$,
the energy decay for the ICS is always slower than for the HCS, see Eq.~(\ref{Haff}). A more explicit form of the generalized Haff's law (\ref{Haffgen}) is
\begin{equation}
p(t)=\left[1+t\sqrt{c+(a-c)\,\frac{\mbox{E}(s)}{\mbox{K}(s)}}\right]^{-2}. \label{Haffgen2}
\end{equation}

Now we consider the particular case of the single hole solution $w_0(m)$.  As $\langle w_0(m) \rangle
\simeq \sqrt{6/{\cal L}}$, we obtain for the pressure (in the physical units)
\begin{equation}\label{pressurehole}
p(\tau) = p_0 \exp \left( -2\sqrt{6} \gamma {\cal L}^{-1/2} \tau
\right)\,.
\end{equation}
Using Eq.~(\ref{t}), we find the original (physical) time $t$ in terms of $\tau$ (again, in the physical units):
\begin{equation}\label{ortimehole}
 t=\frac{(6 {\cal L})^{1/2}}{3 \Lambda \rho_0^{1/2} p_0^{1/2}}
 \left(e^{\sqrt{6} \gamma {\cal L}^{-1/2}
 \tau}-1\right)\,.
\end{equation}
This yields a generalized Haff's law
\begin{equation}\label{Haff1}
    p(t)=\frac{p_0}{\left(1+t/\tilde{t}_c\right)^2}
\end{equation}
with a characteristic cooling time
\begin{equation}\label{newcoolingtime}
\tilde{t}_c=\frac{(6{\cal L})^{1/2}}{3\Lambda \rho_0^{1/2} p_0^{1/2}}\,.
\end{equation}
As ${\cal L}\gg 1$, the cooling time $\tilde{t}_c$ is much longer than
the cooling time $t_c$ corresponding to the HCS:
\begin{equation}\label{compare2}
    \frac{t_c}{\tilde{t}_c}=\left(\frac{6}{{\cal L}}\right)^{1/2} \ll 1.
\end{equation}

\subsection{Relaxation to the single hole state}
 Here we study the late-time dynamics of relaxation of the cooling gas towards the single hole state: the cooling state observed for  ${\cal L} \gg 1$, that is, for $l_{cr}\ll L \ll l_s$. We put $w(m,\tau)=w_0(m)+w_1(m,\tau)$, where $w_0(m)$ is the single hole
asymptotics (\ref{cosh1}), and linearize Eq.~(\ref{eq5}) with respect to the
small correction $w_1$. We obtain
\begin{equation}\label{linear1}
    w_0\frac{\partial w_1}{\partial \tau}=\left(2 w_0\left\langle w_0 \right\rangle -1
    \right)w_1+\left\langle w_1\right\rangle w_0^2 + \frac{\partial^2 w_1}{\partial m^2}\,.
\end{equation}
In the language of the linear stability analysis, the conservation law (\ref{conslaw}) becomes $\left\langle w_0(m)w_1(m,\tau)\right\rangle=0$. Integrating Eq.~(\ref{linear1}) over the box, one can see that, once this condition holds at $\tau=0$, it continues to hold at $\tau>0$.

As will become clear shortly, a natural complete set of eigenfunctions
for the linear equation~(\ref{linear1}) is provided by the following eigenvalue problem:
\begin{equation}\label{eigen}
    y_n^{\prime\prime}(m)+\left[-1+ \lambda_n w_0(m)\right]y_n(m)=0\,.
\end{equation}
for the eigenfunctions $y_n(m)$ obeying the BCs $y_n(\pm \infty)=0$. (Here we have moved the boundaries to infinity which is accurate with an exponential accuracy in the large parameter ${\cal L} \gg 1$.)
Equation~(\ref{eigen}) can be viewed as a stationary Shr\"{o}dinger equation
(with $\hbar=1$) for a particle with
mass $1/2$ and a \emph{fixed} energy $-1$ in the P\"{o}schl-Teller potential well,  see
\textit{e.g.} Ref. \cite{LLQM}. The depth of the well is determined by the eigenvalues $\lambda_n$.
The spectrum of this problem is
discrete:
\begin{equation}\label{eigenvalues}
    \lambda_n=\frac{(n+2) (n+3)}{(6 {\cal L})^{1/2}}\,,\;\;\;\;\;n=0,1,2,3, \dots\,.
\end{equation}
For even values of $n$ one obtains even eigenfunctions:
\begin{eqnarray}
  \nonumber y_n^{even}(m)&=& A_n\cosh ^{2 n+3}   \left(\frac{m}{2}\right) \\
 &\times&_2F_1\left[n+\frac{1}{2},n+\frac{5}{2};\frac{1}{2};-\sinh
   ^2\left(\frac{m}{2}\right)\right],  \label{funceven}
\end{eqnarray}
whereas for odd values of $n$ one obtains odd eigenfunctions:
\begin{eqnarray}
   \nonumber y_n^{odd}(m)&=& B_n\cosh ^{2 n+4}\left(\frac{m}{2}\right) \sinh \left(\frac{m}{2}\right)\\
   &\times& _2F_1\left[n+\frac{3}{2},n+\frac{7}{2};\frac{3}{2};-\sinh
   ^2\left(\frac{m}{2}\right)\right]. \label{funcodd}
\end{eqnarray}
Here $_2F_1(\dots)$ is the hypergeometric function, and $A_n$ and $B_n$ are constants
that we fix using the orthonormality conditions
\begin{equation}\label{orthonorm}
    \int_{-\infty}^{+\infty}y_k(m) y_n(m) w_0(m)\,dm=\delta_{kn}\,,
\end{equation}
the Kroneker delta. The fundamental mode $y_0(m)$ is even, it is proportional to $w_0(m)$:
$$y_0(m)=C_0 \,w_0(m)=(75/2{\cal L})^{1/4}\,\cosh^{-2}(m/2)\,,$$
where $C_0=5^{1/2}6^{-1/4} {\cal L}^{-3/4}$.
The next mode is the first odd eigenfunction $y_1(m)$,  proportional to $dw_0(m)/dm$:
$$
y_1(m)= - \frac{3^{1/4} \sqrt{35} \cosh^{-2}\left(m/2\right) \tanh \left(m/2\right)}
{2^{7/4} {\cal L}^{1/4}}\,.
$$
The next one is the second even eigenfunction
$$
y_2(m)= -\frac{3^{3/4} \sqrt{5} \,(3 \cosh m-4) \cosh^{-4}(m/2)}{2^{7/4} {\cal L}^{1/4}}\,,
$$
and so on. Let us expand $w_1(m,\tau)$ in this complete set of eigenfunctions:
$$
w_1(m,\tau)=\sum_{n=0}^{\infty} a_n(\tau)\, y_n(m)\,,
$$
substitute this expansion in Eq.~(\ref{linear1}), multiply the resulting equation by $y_k(m)$, $k=0,1,2,\dots$ and integrate over $m$ from $-\infty$ to $\infty$. Using Eq.~(\ref{eigen}), we arrive at the following equations for the time-dependent amplitudes $a_k(\tau)$:
\begin{equation}\label{nonzeromodes}
    \frac{d a_k(\tau)}{d \tau}=-\Gamma_k a_k(\tau)\;\;\;\mbox{for}\;k\neq 0\,,
\end{equation}
and
\begin{equation}\label{zeromode}
    \frac{d a_0(\tau)}{d \tau}=2\left\langle w_0\right\rangle \,a_0(\tau) +\frac{1}{C_0} \sum_{n=1}^{\infty} a_{2n}(\tau) \left\langle y_{2n}\right\rangle\,.
\end{equation}
Here
\begin{equation}\label{damping1}
    \Gamma_k=\lambda_k-2\left\langle w_0\right\rangle=\frac{(k-1)(k+6)}{(6 {\cal L})^{1/2}}\,,\;\;\;k=1,2,\dots\,,
\end{equation}
and we have used the equality $ \lambda_0=\langle w_0 \rangle$. The amplitude
equations (\ref{nonzeromodes}) and (\ref{zeromode}), together with the initial
conditions $a_k(0)$, $k=0,1,2, \dots$, enable us to solve the initial value
problem for the evolution of the small perturbation $w_1(m,\tau)$.
Equations~(\ref{nonzeromodes}) show that each of the odd and even modes
$k=1,2,3, \dots$ evolve independently of other modes: the $k=1$ mode has a zero
decay rate (which is expected, as it is a translational mode), while the higher
modes decay exponentially in time $\tau$ :
 \begin{equation}\label{others}
    a_k(\tau)=a_k(0) \exp(-\Gamma_k \tau)\,, \;\;\;k=1,2,3,\dots .
 \end{equation}
The $k=0$ mode behaves quite differently from other modes, as it is affected by
the rest of the \emph{even} modes of the system, see Eq.~(\ref{zeromode}). The
solution of Eq.~(\ref{zeromode}) is:
\begin{eqnarray}
\nonumber  a_0(\tau) &=& \left[a_0(0)+\frac{1}{C_0} \sum_{n=1}^{\infty} \frac{a_{2n}(0)\left\langle y_{2n}\right\rangle}{\lambda_{2n}}\right]\, \exp \left(2\left\langle w_0 \right\rangle \tau \right) \\
  &-& \frac{1}{C_0}\sum_{n=1}^{\infty} \frac{a_{2n}(0)\left\langle y_{2n}\right\rangle}{\lambda_{2n}} \,
  \exp\left(-\Gamma_{2n} \tau\right)\,.
  \label{a0}
\end{eqnarray}
Now we prove that the term in the square brackets vanishes. At $\tau=0$ the conservation law (\ref{conslaw})
can be written as
$$
\left\langle w_0(m) \, \sum_{n=0}^{\infty}a_{2n}(0)y_{2n}(m)\right\rangle = 0,
$$
which yields
\begin{equation}\label{zero}
    a_0(0) + \frac{1}{C_0}\sum_{n=0}^{\infty} a_{2n}(0)\left\langle w_0 y_{2n} \right\rangle=0\,.
\end{equation}
By virtue of the identity $\left\langle y_{2n}\right\rangle=\lambda_{2n} \left\langle w_0 y_{2n}\right \rangle$
[which readily follows from Eq.~(\ref{eigen})], the left side of Eq.~(\ref{zero}) coincides
with the term in the square brackets in Eq.~(\ref{a0}). Therefore, the final result for $a_0(\tau)$
is
\begin{equation}\label{a0final}
    a_0(\tau) = - \frac{1}{C_0}\sum_{n=1}^{\infty} \frac{a_{2n}(0)\left\langle y_{2n}\right \rangle}{\lambda_{2n}} \,
  \exp\left(-\Gamma_{2n} \tau\right)\,.
\end{equation}
$a_0(\tau)$ can behave non-monotonically at short times. However, it always decays at long times, and the dominant decay rate,  at  $\tau \gg {\cal L}^{1/2}$, is $\Gamma_2$.

\begin{figure}[ht]
\includegraphics[scale=1.8]{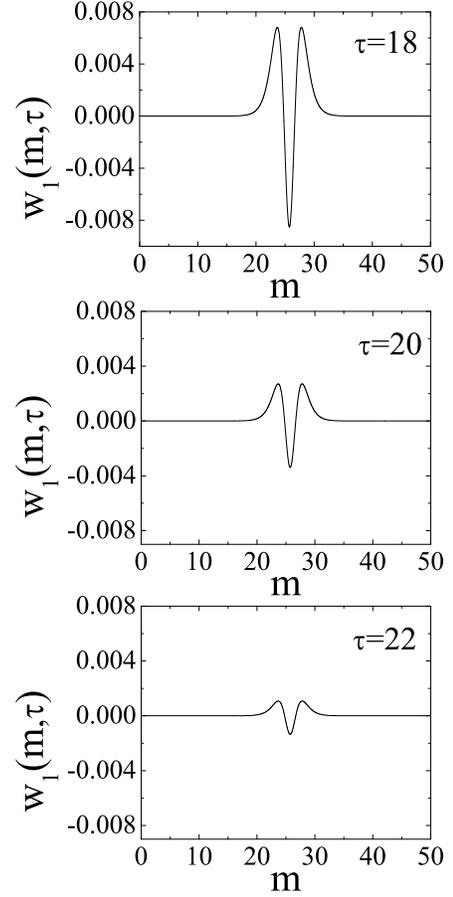}
\caption{The difference $w_1(m,\tau)$ between the time-dependent solution and
the single-hole steady state asymptotics (\ref{cosh1}) at different (late) times for the simulation shown in Fig. \ref{snapshots1}.} \label{dampshape} 
\end{figure}

\begin{figure}[ht]
\includegraphics[scale=0.60]{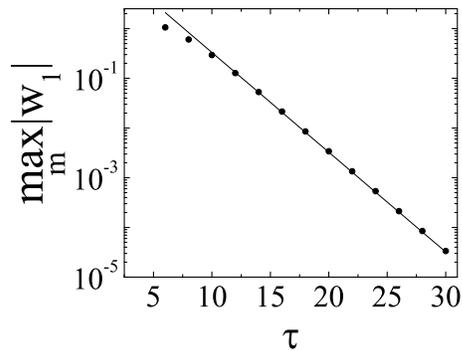}
\caption{Testing the linear stability analysis of the single hole solution. The circles show, in the logarithmic scale, $\mbox{max}_m \,|w_1(m,\tau)|$ (see Fig.~\ref{dampshape}) versus time $\tau$. The solid line depicts our theoretical prediction for long times, when the relaxation is
dominated by the mode $y_2$, so that $\mbox{max}_m \,|w_1|=c_{0}\exp(-\Gamma_{2}\tau)$, where $\Gamma_{2}=8/\sqrt{6{\cal L}}\simeq 0.462$. The adjustable parameter $c_{0}=33.5$.} \label{dampexp} 
\end{figure}

Figures \ref{dampshape} and \ref{dampexp} present a comparison of the linear stability analysis with the simulation shown in Fig. \ref{snapshots1}. Figure \ref{dampshape} shows, at late times, the deviation of the numerical solution from the theoretical single-hole steady state asymptotics (\ref{cosh1}) for the simulation shown in Fig. \ref{snapshots1}. As time proceeds, the deviation tends to zero as expected. Figure \ref{dampexp} compares the numerically observed decay rate of the deviation with the analytical result (\ref{damping1}) for the decay rate $\Gamma_2$ that dominates at late times, and very good agreement is observed.

Using Eqs.~(\ref{ortimehole}) and (\ref{damping1}), we can see that the exponential decay in $\tau$ of each of the eigenmodes $k=1,2, \dots$, see Eq.~(\ref{others}),  becomes a power-law decay
in the physical time:
$$
 a_k(t)=a_k(0) \left(1+\frac{t}{\tilde{t}_c}\right)^{-\frac{(k-1)(k+6)}{6 \gamma}}\,,\;\;\;k=1,2,\dots\,,
$$
with $\tilde{t}_c$ from Eq.~(\ref{newcoolingtime}). The zero mode dynamics (\ref{a0final}) can be represented as a superposition of terms, each of which decaying as a power law in the physical time. Therefore, the mismatch
$w(m,t)-w_0(m)$ between the time-dependent solution $w(m,t)$ and the single hole solution $w_0(m)$ decays, at long times, as $\sim (t/\tilde{t}_c)^{-4/(3\gamma)}$.

Before concluding this section we note that the $k=1$ mode turned out to be marginally stable
because we neglected corrections exponentially small with respect to the
rescaled system length ${\cal L}$.  In a more accurate treatment this mode would
cease to be a translational mode and acquire a non-zero (although exponentially
small) damping rate in time $\tau$. This would lead to a power law decay of this
mode in time $t$ with a power exponent that is exponentially small in ${\cal
L}$.

\section{Coarsening dynamics and statistics of holes}
\label{coarsening}

\begin{figure}[ht]
\includegraphics[scale=0.77]{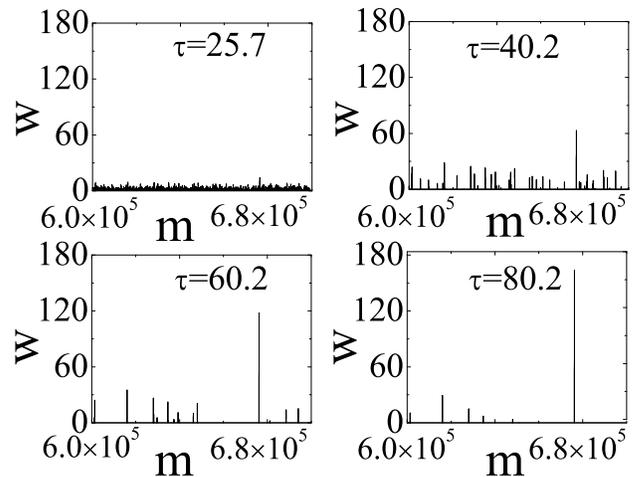}
\caption{Nucleation and coarsening of holes when starting from a small amplitude ``white noise" density perturbation around $w=1$.
Shown is a small fragment of the system of rescaled length/mass
${\cal L}=10^6$ at indicated times.}
\label{M1e6d}
\end{figure}

Numerical simulations with Eq.~(\ref{eq5}) show that, for a sufficiently large rescaled length/mass of the system,  ${\cal L}\gg 1$, \textit{many} peaks of $w$ (hence, holes of the gas density) nucleate in the system \cite{nottoolong}. The nucleation stage, as observed numerically, is shown in the upper left panel of Fig.~\ref{M1e6d}. The initial condition $w(m,\tau=0)$ simulated white noise, as we chose
$w^2(m,\tau=0)$ to be equal to $1$ plus a sum of a very large number of Fourier harmonics with (very small) random amplitudes drawn from a uniform distribution. As evidenced by Fig.~\ref{M1e6d}, the further evolution of the holes resembles
Ostwald ripening \cite{Ostwald}. At this stage nucleation of new holes does not
occur anymore, and a competition between the holes begins.  Underdense holes
release their material into the environment and become more pronounced (even less
dense), while holes with more material continue to suck the material in until they
disappear. At some stage the holes which gas density previously decreased,
reverse the trend and begin to densify. At the end of this coarsening process
only one hole (that was the least dense in the beginning) remains and forms the
single-hole solution (\ref{cosh1}) and (\ref{cosh2}) \cite{switch}. Clearly, the holes compete
non-locally: via the spatial averaging term of Eq.~(\ref{eq5}).

Can one build upon the analogy with Ostwald ripening and develop an
asymptotic theory of the hole coarsening dynamics? Consider a late stage of the dynamics when there are $N$ holes, located sufficiently
far from each other, and centered at points $m_i$, $i=1,2, \dots, N$. A simple theory assumes that the spatial shape of each hole coincides with that of the limiting steady
state asymptotics~(\ref{cosh1}), but with its own amplitude $A_i(t)$ that
depends on time. The latter assumption is based on a remarkable fact that, up to exponentially small corrections, Eq.~(\ref{eq5})
admits the following ansatz:
\begin{equation}\label{ansatz}
w(m,\tau)=\sum_{i=1}^N A_i(\tau) \, \cosh^{-2}\left(\frac{m-m_i}{2}\right)\,.
\end{equation}
Plugging it into Eq.~(\ref{eq5}) and neglecting exponentially small overlap
terms, we find that the equation is satisfied once the following $N$ relations
hold:
\begin{equation}\label{Adot}
\dot{A}_i(\tau)=S(\tau) A_i(\tau)-\frac{3}{2}\,,\;\;\;\;\;\;i=1,2, \dots, N\,.
\end{equation}
Here
\begin{equation}\label{S}
S(\tau) =\frac{4}{{\cal L}} \,\sum_{i=1}^N A_i(\tau) \simeq \left\langle w(m,\tau)\right\rangle\,.
\end{equation}
Once all the initial amplitudes $A_i(0)$ of the holes are known, the effective dynamical
system~(\ref{Adot}) provides a complete description of the problem. The conservation law (\ref{conslaw}) of the original Eq.~(\ref{eq5}) becomes an integral
of motion of the dynamical system~(\ref{Adot}):
\begin{equation}\label{integral1}
\sum_{i=1}^N A_i^2(\tau) = \frac{3 {\cal L}}{8} =const\,.
\end{equation}
Equations~(\ref{Adot})-(\ref{integral1}) are similar to (the discrete version of)
the Lifshitz-Slyozov theory of Ostwald ripening \cite{LS},
and their properties give a \textit{qualitative} explanation to the
properties of coarsening observed in Fig.~\ref{M1e6d}. Indeed, the holes with amplitudes
greater than the (time-dependent) critical amplitude
$A_{cr}(\tau)=(3/2)S^{-1}(\tau)$ grow in the amplitude, while holes with amplitudes less than
$A_{cr}(\tau)$ decrease their amplitude and disappear. As $A_{cr}(\tau)$ grows with time, the holes that previously grew in the amplitude begin to decrease their amplitude and finally disappear.

\begin{figure}[ht]
\includegraphics[scale=1.2]{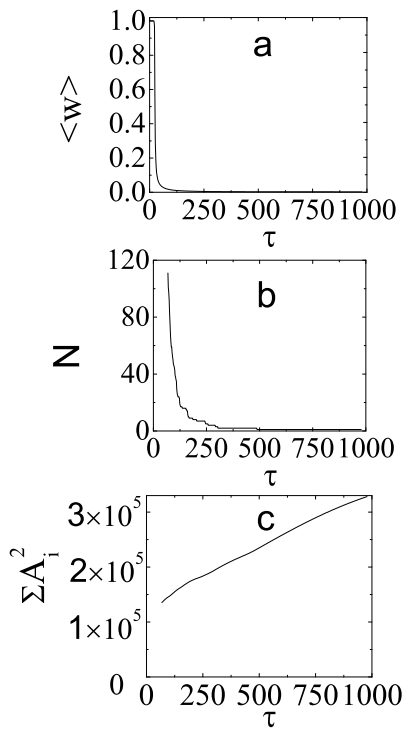}
\caption{The time histories of $\langle w \rangle$ (a), the number of holes $N$ (b) and the sum of the hole amplitudes squared (c) for the ``generic" simulation (starting from a small amplitude noise) shown in Fig.~\ref{M1e6d}.} \label{M1e6e}
\end{figure}

A natural further step is to assume $N\gg 1$, treat the hole amplitude as a continuous variable and deal with the probability distribution $F(A,\tau)$ of
the hole amplitudes $A$ at time $\tau$.  The corresponding theory can be formulated in the spirit of the Lifshitz-Slyozov theory of Ostwald ripening, and we present it in Appendix B. How does this theory compare
with numerical simulations? Figure~\ref{M1e6e} presents some quantitative
characterization of the hole coarsening dynamics for the numerical simulation shown in
Fig.~\ref{M1e6d}. Shown are the time histories of $\langle w \rangle$ (panel a), of
the total number of holes in the system $N$ (panel b) and of the
sum of the hole amplitudes squared (panel c) for the simulation
shown in Fig.~\ref{M1e6d}.  [Because of the noisy initial condition, it
takes some time for well-defined holes to nucleate. We started the hole
count at the time when the total number of the local maxima of $w(m)$
became equal, for the first time, to the total number of $m$-intervals
where $w$ was less than a prescribed small threshold $10^{-4}$.] One
can immediately see on the lower panel of Fig.~\ref{M1e6e} that the
conservation law (\ref{integral1}) is \textit{not} obeyed in this simulation.
It is not surprising, therefore, that other quantitative predictions of our
Lifshitz-Slyozov-type theory, see Appendix B, are also \textit{not} supported by this simulation.
Most directly, the shape of an individual hole does \textit{not} agree with that
assumed in the ansatz (\ref{ansatz}). The holes observed in this ``generic"
simulation have a more complicated structure, and are not characterizable by
a single parameter such as $A_i(\tau)$.

\begin{figure}[ht]
\includegraphics[scale=0.70]{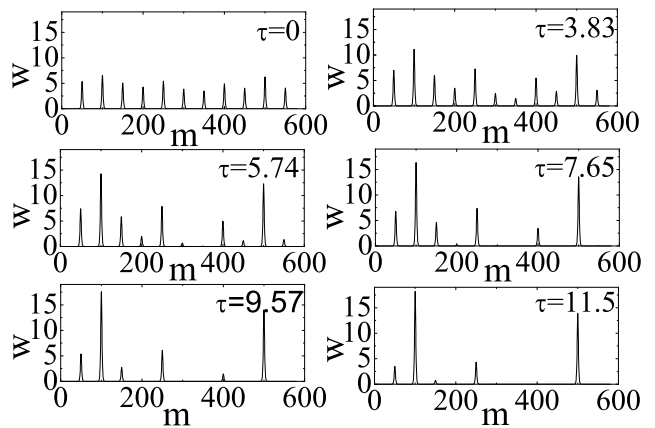}
\caption{Coarsening of holes when starting from the ansatz (\ref{ansatz}) with $N_0=2\times 10^4$ holes. The initial hole amplitudes $A_i$ are randomly distributed according to a (positive) half-gaussian with variance 1. This distribution is normalized by the condition $\sum_{i=1}^{N_0}A_{i}^{2}=3{\cal L}/8$. Shown is a small fragment of the system of rescaled length/mass ${\cal L}=10^6$.}\label{M1e6a}
\end{figure}

\begin{figure}[ht]
\includegraphics[scale=1.4]{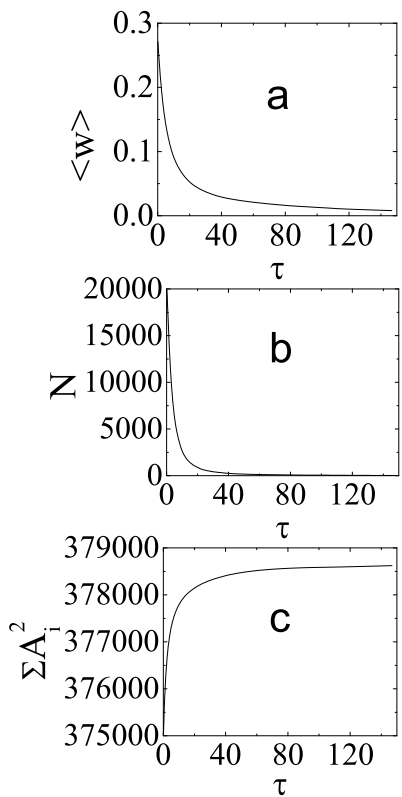}
\caption{The time histories of $\langle w \rangle$ (a), the number of holes $N$ (b) and the sum of the hole amplitudes squared (c) for the simulation that started from the ansatz (\ref{ansatz}) and is shown in Fig.~\ref{M1e6a}. Theoretical prediction (\ref{integral1}) for $\sum A_i^2$ is $3{\cal L}/8=375,000$ which agrees with the simulation within a 1\% error.} \label{M1e6b}
\end{figure}

\begin{figure}[ht]
\includegraphics[scale=1.1]{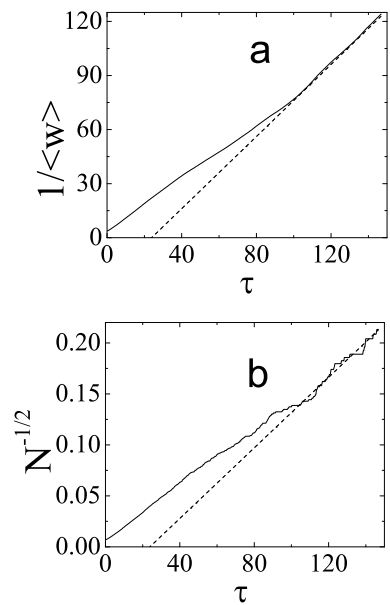}
\caption{A comparison of the time histories of $\langle w \rangle$ and $N$ from Fig.~ \ref{M1e6b} with theoretical predictions. The solid line in panel a shows the numerical results for
$1/\langle w \rangle$ versus time $\tau$.  The dashed line shows our theoretical prediction for late times: $\langle w\rangle=1/(\tau-\tau_f)$, where $\tau_f$ is an adjustable parameter (in this simulation $\tau_f\simeq 23.9$).   Plotted in panel b is the numerical result for $N^{-1/2}$ versus $\tau$ (the solid line), and the theoretical prediction
$N^{-1/2}(\tau)= \sqrt{3/{\cal L}}\,(\tau-\tau_f)$ with no additional adjustable parameters (the dashed line).
Here $\sqrt{3/{\cal L}}\simeq 1.73 \times 10^{-3}$.  The noise, evident in panel b at late times, is due to a small number of holes at those times.}\label{M1e6c}
\end{figure}

It is therefore remarkable, that the ansatz (\ref{ansatz}) does describe a \textit{stable} regime of coarsening. That is, if one starts the simulation, at $\tau=0$, with an ensemble of holes with different amplitudes, describable by the ansatz (\ref{ansatz}), the ansatz continues to hold and, moreover, the system approaches the simple scaling regime
predicted by our theory of Lifshitz-Slyozov type.   The results of one such simulation are presented in Figs.~\ref{M1e6a} - \ref{M1e6c}. Here the holes were placed at a (sufficiently large) equal distance from each other, and the initial hole amplitudes $A_i$ were chosen randomly from a positive half-gaussian with variance 1.  One can see a hole coarsening process in Fig.~\ref{M1e6a}: holes with a larger amplitude (that is, with less gas) grow (that is, loose gas) at the expense of holes with a smaller amplitude. The time histories of $\langle w \rangle$  and the number of holes $N$, presented in Fig.~\ref{M1e6b}, resemble those for the previously described ``generic" simulation. The behavior of the sum $\sum_1^N A_i^2$ is, however, dramatically different: here the conservation law (\ref{integral1}) is obeyed with a $1$ percent accuracy. A closer inspection of the time histories of $\langle w\rangle$ and $N(\tau)$ (see Fig.~\ref{M1e6c}) shows that, at late times, these quantities agree with the theoretical predictions from Eqs.~(\ref{mus}) (with $\mu_1=1$)
and (\ref{averagesspecial}), presented in Appendix B. Indeed, by using only one adjustable parameter: the time shift $\tau_f$, related to the time  of approaching the scaling regime, we obtained good agreement for the two different quantities. We also checked (not shown) that, at different times, the shapes of individual holes are very well described by the $\cosh^{-2}$ profile assumed in the ansatz (\ref{ansatz}).

\section{Summary and Discussion}
\label{summary} We have developed a nonlinear theory of low Mach number channel flows of
freely cooling dilute granular gases with nearly elastic particle collisions. We focused on the
case when the sound travel time through the system is much shorter than the
cooling time and the heat diffusion time. Then, after a brief transient, the gas
pressure becomes (almost) uniform in space. This makes it possible to reduce the
granular hydrodynamic equations,  in Lagrangian coordinates, to a single
nonlinear and nonlocal equation of a reaction-diffusion type. With heat
diffusion neglected, the reduced equation becomes integrable, and any
inhomogeneous initial condition  produces a finite-time density blowup. The
density blowup has the same universal features at singularity as those exhibited
by a family of exact solutions of the full set of ideal hydrodynamic equations
\cite{Fouxon1,Fouxon2}. The density blowup, however, is arrested by the
heat diffusion. As a result a novel, inhomogeneous cooling state (ICS) of the
gas emerges which has a time-independent density profile.  For channels of an intermediate length that we considered, the ICS represents a global
attractor of the system. Both its structure, and the late-time relaxation towards it are determined by a single
dimensionless parameter ${\cal L}$ which is of the order of the ratio of the channel
length to the critical length predicted by the linear theory of instability of the homogeneous cooling state.  The energy decay of the ICS differs considerably from Haff's law: the characteristic decay time diverges with the size of the system as ${\cal L}^{1/2}$, see Eq.~(\ref{compare2}). At large ${\cal L}$, the maximum density of the ICS grows exponentially with ${\cal L}$. Therefore, for sufficiently long channels (the rest of parameters being fixed), the dilute gas assumption breaks down, and close packed regions emerge.

For ${\cal L} \gg 1$ the cooling
dynamics proceeds as a competition between ``holes". This competition is
quite similar to Ostwald ripening. In the simple case when the initial
state consists of $N$ well separated holes $\sim \cosh^{-2}(m/2)$, the analogy
with Ostwald ripening becomes complete, as the ``hole ripening" statistics
exhibits a simple dynamic scaling behavior and is describable by a variant of
the Lifshitz-Slyozov theory. Here, in analogy with other phase ordering systems with a conserved order parameter,
the probability distribution of the holes with respect to their
amplitudes approaches, at long times, the special (limiting) self-similar solution, that
is analytic at the edge of its (compact) support.   However, for a generic, noisy initial condition,
the competing holes have a more complicated structure than that described by the ansatz
(\ref{ansatz}). This brings about a lack of simple dynamic scaling.  A theory of this regime
has yet to be developed.

In the light of the above results, a non-linear development of the clustering instability of the HCS, for intermediate channel lengths, is but a particular case of our low Mach number theory. Ultimately, the instability transforms an (almost) homogeneous initial gas density profile into an inhomogeneous but time-independent density profile: the ICS described above. For ${\cal L} \gg 1$  this transformation occurs through an intermediate state with many holes (and many clusters).

It would be interesting to investigate the ICSs, and relaxation toward them, in
MD simulations. To directly test our low Mach number theory, one should choose the MD simulation
parameters so as to guarantee the length scale separation $l_{cr} < L \ll l_s$ assumed here. We stress that this hierarchy of length scales demands nearly elastic particle collisions: $\sqrt{1-r^2}\ll 1$.  In addition,
the channel length $L$ should not be too large so that the theoretically
predicted maximum gas density in the ICs is still small compared to the close packing
density of spheres.

It is worth noticing that, in all asymptotic cooling regimes of an inhomogeneous gas that we have investigated, the energy decays slower than in the case of a HCS.  Haff's cooling law, therefore, provides an upper bound on the energy decay rate. In fact, this is a general theorem, universally valid for a low Mach number flow. Indeed, according to Eq.~(\ref{pressure0}), the logarithmic derivative of the pressure (and, therefore, of the total energy) is proportional to $- \langle w \rangle$. For a HCS $\langle w \rangle=1$, whereas for \textit{any} ICS $\langle w \rangle < 1$, by virtue of the Cauchy-Schwarz inequality and the identity $\langle w^2 \rangle=1$.

What can be said about the opposite, long-wavelength limit, $\lambda \gg l_s$, where
$\lambda$ is the characteristic length scale of the initial perturbations?
Although there has been some progress in this case \cite{ELM,MP,Fouxon1}, a
complete understanding of the dynamics and structure of the flow is still
lacking. It should be possible to derive a
different reduced model in that limit, and see
whether the popular ``pressure instability scenario"
\cite{Goldhirsch} is at work there. (It is clear that the pressure instability scenario is irrelevant in
the intermediate wavelength limit, considered in the present paper.)

Note that the ICSs, that we have discovered here, are exact solutions of the \textit{full} set of
granular hydrodynamic equations~(\ref{hydrodynamics1})-(\ref{hydrodynamics3}) for a nearly elastic dilute gas, without any reductions. Therefore, a
question arises on whether the ICS represents an attractor in the general case,
including the long wavelength limit. A complete (unreduced) linear stability
analysis around the ``hole" asymptotics (\ref{cosh1}) could be the first step in an
attempt to answer this question. Such an analysis can be complemented by
numerical hydrodynamic simulations of nonlinear cooling flows, so as to
elucidate possible effects of shock waves on the (nonlinear) stability of the
ICS.

Does this work, limited to channel flows, have any relevance to the shearing/clustering
instability of a freely cooling granular gas in fully multi-dimensional geometries? To begin with, the low Mach number theory can be
extended to the higher dimensions, once the characteristic sound travel distance $l_s$
is much larger than \textit{all} system dimensions. This extension should take into a proper account
the flow vorticity, in much the same way as it was done in Ref. \cite{Glasner}
where a two-dimensional low Mach number flow of an ideal gas, driven by the heat diffusion, was investigated. Although
not very simple, such a reduced description (with the acoustic modes eliminated) will be advantageous compared to the full set of multi-dimensional hydrodynamic equations. Furthermore, the novel ICSs of
the granular gas (that represent exact solutions of the unreduced granular hydrodynamic equations) may have multi-dimensional analogs. Finding these analogs, and investigating their stability with respect to multi-dimensional perturbations which have both potential, and solenoidal velocity components, can be a natural next step in developing a more complete nonlinear theory of the shearing/clustering instability. The channel flow theory developed here (see also Refs. \cite{ELM,MP,Fouxon1,Fouxon2,Puglisi}) sets the ground for the future work.

\begin{acknowledgments}
Our work was supported by the Israel Science
Foundation (grant No. 107/05) and by the German-Israel Foundation
for Scientific Research and Development (Grant I-795-166.10/2003).
\end{acknowledgments}

\section*{Appendix A. Numerical scheme}
\renewcommand{\theequation}{A\arabic{equation}}
\setcounter{equation}{0}  

We employed the following implicit finite difference scheme for a numerical solution of Eq.~(\ref{eq5}):
\begin{equation}\label{discrete}
    \frac{w_{i}^{2}-\hat{w}^{2}_{i}}{2\delta\tau}=
-w_{i}+w_{i}^{2}\frac{\sum_{i=1}^{n}w_{i}}{n}+Dw_{i}\,,
\end{equation}
where
$\delta \tau$ is the time step, $w_{i}=w(m_{i},\tau+\delta\tau)$, $\hat{w}_{i}=w(m_{i},\tau)$. A standard discretization $Dw_{i}$ of the diffusion term was used: for the periodic BCs we put
$$
Dw_{i}=
\left\{\begin{array}{ll}
\frac{w_{2}-2w_{1}+w_{n}}{h^{2}}\,, \qquad\quad i=1\,,\\
\frac{w_{i+1}-2w_{i}+w_{i-1}}{h^{2}}\,, \quad 1<i<n\,, \\
\frac{w_{1}-2w_{n}+w_{n-1}}{h^{2}}\,, \qquad i=n\,.
\end{array}
\right.
$$
where $h={\cal L}/n$ is the grid size. The approximation error of this scheme is ${\cal O}(\delta\tau^{2})$ in $\delta\tau$ and ${\cal O}(h^{3})$ in $h$. Note that the scheme conserves \textit{exactly} the discrete version of the conservation law (\ref{conslaw}), $\langle w_i(\tau)^2\rangle = (1/n)\sum_{i=1}^{n}w_{i}^{2}=1$, once  $\langle w_i(0)^2\rangle = 1$.

We solved the set of nonlinear algebraic equations (\ref{discrete}) by an iteration procedure  based on Newton's method. To obtain, after linearization, a standard cyclic tridiagonal system, we used the values of $w_i$, entering the sum $\sum_i^n w_i$, from the previous iteration.  We demanded that the residual (the maximum of the absolute value of the difference between the left and right hand sides of the equations after the iteration process) be less then $10^{-13}$. Because of the finite residual, this procedure conserved the mean square of $w$ with an almost machine precision, but not exactly.  Therefore, we enforced an even stricter conservation by adding, at each time step, a constant $c$ to the numerical solution $w_i$ found with the iteration procedure. The value of $c$ is determined as follows. We represent the (yet unknown) corrected values $\bar{w}_{i}$ as $\bar{w}_{i}=w_{i}+c$. Then
$$
\langle\bar{w}_{i}^{2}\rangle=\frac{1}{n}\sum_{i=1}^{n}\left(w_{i}^{2}+2cw_{i}+c^{2}\right)=\langle w^{2}\rangle+2c \langle w \rangle+c^{2}\,.
$$
Now we demand that the right hand side be equal to 1. Neglecting the $c^2$ term, we find
$$
c=\frac{1-\langle w^2\rangle}{2\langle w \rangle}\,.
$$
We always obtained $|c|<10^{-14}$ in our computations. This justifies neglecting the $c^2$ term.

The typical set of parameters for the investigation of relaxation towards a stationary single hole asymptotics (\ref{cosh1}) was ${\cal L}=50$ and $n=2.5 \times 10^4$, so $h=2 \times 10^{-3}$. In the hole coarsening simulations we used ${\cal L}=10^6$ and $n=2.8 \times 10^6$, so $h\simeq 0.36$. In all cases the time step was chosen to be $\delta \tau=h^2$.

\section*{Appendix B. Hole coarsening in the spirit of the Lifshitz-Slyozov theory}
\renewcommand{\theequation}{B\arabic{equation}}
\setcounter{equation}{0}  

Here we treat the hole amplitude (see Section \ref{coarsening}) as a continuous variable and deal with the probability distribution $F(A,\tau)$ of
the hole amplitudes $A$ at time $\tau$.  The total number of holes $N(\tau)=
\int_0^{\infty} F(A,\tau)\,dA \gg 1$. As there is no nucleation of new holes and no
hole mergers, the evolution of $F(A,\tau)$ is described, in the spirit of
the Lifshitz-Slyozov theory \cite{LS}, by a continuity equation in the space of
hole amplitudes:
\begin{equation}\label{continuity}
\frac{\partial F}{\partial \tau}+\frac{\partial}{\partial A}\left[\left(S
A-\frac{3}{2} \right)F\right]  =0\,,
\end{equation}
where
\begin{equation}\label{Sint}
    S(\tau) =\frac{4}{{\cal L}} \,\int_{0}^{\infty} A F(A,\tau) dA \,,
\end{equation}
and
\begin{equation}\label{integral2}
\int_{0}^{\infty} A^2 F(A,\tau) dA = \frac{3 {\cal L}}{8}=const\,.
\end{equation}
Equations similar to Eqs.~(\ref{continuity})-(\ref{integral2}) have appeared in
the context of the Lifshitz-Slyozov model of Ostwald ripening \cite{LS} and its
analogs for different transport mechanisms \cite{AMS2,W,MS,CP,GMS}.  In those systems one is
usually interested in the question of whether or not the probability distribution $F(A,\tau)$
approaches, at late times, a self-similar shape. A simple power counting in  Eqs.~(\ref{continuity})-(\ref{integral2}) yields
\begin{equation}\label{SS}
F(A,\tau) = \frac{{\cal L}}{4 \,\tau^3}
\,\Phi\left(\frac{A}{\tau}\right)\,,
\end{equation}
where $\Phi(\eta)\ge0$ is the (yet unknown) rescaled distribution, and the
coefficient ${\cal L}/4$ is chosen for convenience. Using
Eqs.~(\ref{Sint}) and (\ref{integral2}), we obtain
\begin{equation}\label{mus}
    S(\tau)=\frac{\mu_1}{\tau} \;\;\;\;\;\mbox{and} \;\;\;\;\;\mu_2=\frac{3}{2}\,,
\end{equation}
respectively. Here $\mu_k$ is the $k$-th moment of the rescaled distribution:
$\mu_k=\int_0^{\infty}\eta^k \Phi(\eta) d\eta$. One can already see that the
total number of holes $N(\tau)$ goes down as $\tau^{-2}$, while both the average
hole amplitude $\bar{A}(\tau)$ and the critical amplitude $A_{cr}(\tau)$ grow
linearly with $\tau$. The pre-factors of these power laws will be determined once
$\Phi(\eta)$ is found. Plugging Eq.~(\ref{SS}) and the first of Eqs.~(\ref{mus})
into Eq.~(\ref{continuity}) we obtain an ordinary differential equation for $\Phi(\eta)$:
\begin{equation}\label{ode}
    \left[(\mu_1-1)\,\eta -\frac{3}{2}\right]
    \frac{d\Phi}{d\eta}+(\mu_1-3)\,\Phi=0\,,
\end{equation}
whose solution is elementary. As in other variants of the LS-theory, we obtain here a
whole \textit{family} of shape functions $\Phi_{\mu_1}(\eta)$, parameterized by the first
moment $\mu_1$. The solutions exist, with finite moments, for $1\le\mu_1<
\infty$. For $\mu_1>1$ the solutions have finite support:
\begin{equation}\label{support}
\Phi_{\mu_1}(\eta)=\left\{\begin{array}{ll}
B_{\mu_1}\,\left[\frac{3}{2}-(\mu_1-1)\,\eta\right]^{\frac{3-\mu_1}{\mu_1-1}}
& \mbox{if $0<\eta<\eta_m$}\,, \\
0 & \mbox{if $\eta>\eta_m$\,,}
\end{array}
\right.
\end{equation}
where $\eta_m=(3/2)(\mu_1-1)^{-1}$. The constant $B_{\mu_1}$ can be
determined from the second of Eqs.~(\ref{mus}) (that plays the role of a
normalization condition):
$$
B_{\mu_1}=2^{\frac{2\mu_1}{\mu_1-1}}\,3^{-\frac{\mu_1+1}{\mu_1-1}}\,\mu_1
(\mu_1+1)\,.
$$
This yields, for $\mu_1>1$,
\begin{eqnarray}
\label{averages}
  N(\tau) = \frac{{\cal L}\mu_1 (1+\mu_1)}{6\,\tau^2}, && \bar{A}(\tau)=
    \frac{3 \Gamma \left(\frac{2}{\mu_1 -1}\right)\, \tau}
    {(\mu_1 -1)^2 \Gamma \left(\frac{2 \mu_1 }{\mu_1 -1}\right)}\,, \nonumber\\
A_{cr}(\tau) &=& (3 \tau)/(2 \mu_1)\,.
\end{eqnarray}
For $1<\mu_1<3$, the solutions
(\ref{support}) vanish at $\eta=\eta_m$, whereas for $\mu_1>3$ they diverge at
$\eta=\eta_m$. As all the moments $\mu_k$ remain finite, the diverging distributions
are legitimate.

\begin{figure}[ht]
\includegraphics[scale=0.4]{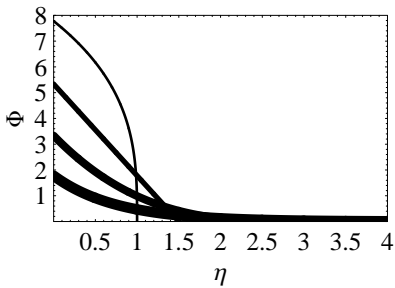}
\caption{Rescaled distributions of the hole amplitudes $\Phi_{\mu_1}(\eta)$ for
$\mu_1=2.5$, $2$, $1.5$, and $1$ (the latter corresponding to the limiting distribution).
Smaller $\mu_1$
are shown by thicker lines.} \label{distributions}
\end{figure}

For $\mu_1=1$ we obtain a \textit{limiting} solution
$\Phi_1(\eta)=(16/9)\,\exp(-4\eta/3)$ that has an infinite support
$0\le\eta<\infty$. The self-similar probability distribution (\ref{SS}) becomes
\begin{equation}\label{SSlimiting}
F(A,\tau) = \frac{4{\cal L}}{9\,\tau^3}
\,\exp\left(-\frac{4A}{3\tau}\right)\,.
\end{equation}
In this case
\begin{equation}\label{averagesspecial}
    N(\tau)=\frac{{\cal L}}{3\,\tau^2},\;\; \bar{A}(\tau)=\frac{3
    \tau}{4},\;\; \mbox{and}\;\; A_{cr}(\tau)=\frac{3 \tau}{2}.
\end{equation}
These expressions also follow from Eqs.~(\ref{averages}) in the limit of
$\mu_1\to 1$.

Figure \ref{distributions} depict the rescaled distributions
$\Phi_{\mu_1}(\eta)$ for four values of the parameter $\mu_1$. Selection of the
``correct" self-similar solution out of the family of solution represents a
subtle problem that was resolved only recently. It turns out that the selection
is only made by (a certain feature of) the initial condition $F(A,\tau=0)$
\cite{AMS2,MS,CP,GMS,Pego}. If $F(A,\tau=0)$ has compact support, the similarity
solution, if any, is selected by the behavior of $F(A,\tau=0)$ near the supremum
$A_{max}$ of the support. If $F(A,\tau=0)$ has a power-law asymptote near
$A_{max}$, the exponent of this power law selects one of the solutions from the
family (\ref{support}). If $F(A,\tau=0)$ goes to zero exponentially fast at
$A\to A_{max}$ (or if the support of $F(A,\tau=0)$ is infinite), the limiting
solution (\ref{SSlimiting}) is selected.

This sensitivity to initial conditions shows a certain lack of robustness of the
Lifshitz-Slyozov model and its analogs like our
Eqs.~(\ref{continuity})-(\ref{integral2}). As a remedy, one has to account for
an additional physics (that may be less universal and more system-dependent).
For example, in the context of the interface-controlled Ostwald ripening  strong
selection is achieved via an account of direct droplet merger events
\cite{CMPS}.

As we show in Section \ref{coarsening}, the  Lifshitz-Slyozov-type model does \textit{not} agree with numerical simulations that start from \textit{generic} initial conditions.
However, if one starts the simulation with an assembly of holes, describable by the ansatz (\ref{ansatz}), the ansatz continues to hold, and the system approaches the simple scaling regime predicted by the Lifshitz-Slyozov-type theory.
Therefore, we want to pursue the ansatz (\ref{ansatz}) a bit further, as it provides an interesting, though non-generic, characterization of the hole coarsening.
We assume that the limiting distribution (\ref{SSlimiting}), corresponding to $\mu_1=1$,
is selected and
use Eq.~(\ref{pressure1}) and the relation $S(\tau)=1/\tau$ to find the corresponding scaling
behavior of the gas pressure $p(\tau)$. We obtain
\begin{equation}\label{pressure5}
    \frac{1}{p(\tau)} \frac{dp}{d\tau}= -2 \gamma S(\tau) =
    -\frac{2\gamma}{\tau}\,,
\end{equation}
which  yields $p(\tau) =p_0 (\tau_0/\tau)^{2\gamma}$, where $\tau_0$ is an
effective ``initial" time, and $p_0=p(\tau_0)$. Using Eq.~(\ref{t}), we find the
following relation between the original (physical) time $t$ and the new time $\tau$:
$$
t =\frac{2 \gamma\,\tau^{\gamma+1}}{(\gamma+1) \Lambda
\rho_0^{1/2}p_0\tau_0^{\gamma}}\,.
$$
As a result,
$$\frac{p(t)}{p_0} = \left[\frac{2 \gamma\,\tau_0}{(\gamma+1) \Lambda
\rho_0^{1/2}p_0\,t}\right]^{\frac{2\gamma}{\gamma+1}} =
\left(\frac{t_0}{t}\right)^{\frac{2\gamma}{\gamma+1}}\,,
$$
where $t_0=t(\tau_0)$. Now, in the low Mach number regime we have been dealing with throughout this paper, the total energy of the gas decays in (almost) the same way as the pressure, so $E_{tot}(t)
\sim t^{-\frac{2\gamma}{\gamma+1}}$. We obtain $E_{tot}(t) \sim t^{-4/3}$ and $E_{tot}(t)
\sim t^{-5/4}$  in 2d (disks) and 3d (spheres),
respectively. Again, the cooling dynamics proceeds slower than that predicted
by Haff's law (\ref{Haff}). We checked that the same conclusion holds for any $\mu_1$, that is for all possible self-similar distributions of the hole amplitudes.


\begin{thebibliography} {99}
\bibitem{BP} N.V. Brilliantov and T. P\"{o}schel,  \textit{Kinetic Theory of Granular
Gases} (Oxford University Press, Oxford, 2004).
\bibitem{Goldhirsch2} I. Goldhirsch,  Annu. Rev. Fluid Mech. \textbf{35},
267 (2003).
\bibitem{Hopkins} M. A. Hopkins and M. Y. Louge, Phys. Fluids A \textbf{3}, 47 (1991).
\bibitem{Goldhirsch} I. Goldhirsch and G. Zanetti, Phys. Rev. Lett. \textbf{70}, 1619 (1993); I. Goldhirsch, M.-L. Tan, and
G. Zanetti,  J. Sci. Comp. \textbf{8}, 1 (1993).
\bibitem{McNamara1} S. McNamara, Phys. Fluids A \textbf{5}, 3056 (1993).
\bibitem{McNamara2} S. McNamara and W. R. Young,  Phys. Rev. E \textbf{53}, 5089 (1996).
\bibitem{Ernst} R. Brito and M. H. Ernst,  Europhys.
Lett. \textbf{43}, 497 (1998).
\bibitem{Brey}  J. J. Brey,  M. J. Ruiz-Montero, and D. Cubero, Phys. Rev. E \textbf{60}, 3150 (1999).
\bibitem{Luding} S. Luding and H. J. Herrmann,  Chaos \textbf{9}, 673 (1999).
\bibitem{van Noije} T.P.C. van Noije and M.H. Ernst, Phys. Rev. E \textbf{61}, 1765 (2000).
\bibitem{Ben-Naim2}  X. B. Nie, E. Ben-Naim, and S. Y. Chen,  Phys. Rev. Lett. \textbf{89}, 204301 (2002).
\bibitem{ELM}  E. Efrati, E. Livne, and B. Meerson,  Phys. Rev. Lett. \textbf{94},
088001 (2005).
\bibitem{MP} B. Meerson and A. Puglisi, Europhys. Lett. \textbf{70}, 478 (2005).
\bibitem{Garzo} V. Garz\'{o}, Phys. Rev. E \textbf{72}, 021106 (2005).
\bibitem{Bromberg} Y. Bromberg, E. Livne, and B. Meerson,  in \textit{Granular Gas
Dynamics}, edited by T. P\"{o}schel and N.V. Brilliantov (Springer, Berlin,
2003), p. 251; cond-mat/0305557.
\bibitem{Volfson} D. Volfson, B. Meerson, and L. S. Tsimring,  Phys. Rev. E  \textbf{73}, 061305 (2006).
\bibitem{Fouxon1} I. Fouxon, B. Meerson,  M. Assaf, and E. Livne, Phys.
Rev. E \textbf{75}, 050301(R) (2007).
\bibitem{Fouxon2} I. Fouxon, B. Meerson,  M. Assaf, and E. Livne, Phys.
Fluids \textbf{19}, 093303 (2007).
\bibitem{Whitham} G.B. Whitham, \textit{Linear and Nonlinear Waves}
(Wiley, New York, 1974), Chapter 2.
\bibitem{Puglisi} A. Puglisi, M. Assaf, I. Fouxon, and B. Meerson, Phys. Rev. E (in press).
\bibitem{Zeldovich} A.G. Doroshkevich and Ya. B. Zel'dovich, Zh. Eksp. Teor. Fiz.
\textbf{80}, 801 (1981) [Sov. Phys. - JETP 53, 405 (1981)].
\bibitem{Meerson89a} B. Meerson, Phys. Fluids A \textbf{1}, 887 (1989).
\bibitem{Meerson89b} B. Meerson, Astrophys. J. \textbf{347}, 1012 (1989).
\bibitem{AMS1} I. Aranson, B. Meerson, and P.V. Sasorov, Phys. Rev. E \textbf{47}, 4337 (1993).
\bibitem{AMS2} I. Aranson, B. Meerson, and P.V. Sasorov, Phys. Rev. E \textbf{52}, 948 (1995).
\bibitem{Kaganovich} D. Kaganovich, B. Meerson, A. Zigler, C. Cohen, and
J. Levin, Phys. Plasmas \textbf{3}, 632 (1996).
\bibitem{Glasner} A. Glasner, E. Livne, and B. Meerson, Phys. Rev. Lett. \textbf{78}, 2112 (1997).
\bibitem{MeersonRMP} B. Meerson, Rev. Mod. Phys. \textbf{68}, 215 (1996).
\bibitem{ZR} Ya. B. Zel'dovich  and Yu. P. Raizer,  \textit{Physics of Shock Waves and
High Temperature Hydrodynamic Phenomena, Vol. 1} (Academic Press, New York,
1966).
\bibitem{Abramowitz} M. Abramowitz, \textit{Handbook of
Mathematical Functions} (National Bureau of Standards, Washington, 1964).
\bibitem{nottoolong} On the other hand, we assume  throughout this paper that the channel is not
too long, so that the uniform pressure approximation remains valid. Long
channels imply, in the low Mach number theory, the double inequality $l_{cr}\ll
L\ll l_s$. In terms of the rescaled length/mass of the system ${\cal L}$, long
channels imply $1\ll {\cal L} \ll (1-r^2)^{-1/2}$.
\bibitem{switch} This scenario assumes periodic BCs.
For the no-flux  BCs one finally obtains  one-half of the hole,
with the density minimum and maximum at the channel ends.
\bibitem{LLQM} L.D. Landau and E.M. Lifshitz, \textit{Quantum Mechanics. Non-Relativistic Theory}
(Pergamon, London, 1965), p. 72.
\bibitem{Ostwald} W. Ostwald, Z. Phys. Chem., Stoechiom. Verwandtschaftsl. \textbf{34},
495 (1900).
\bibitem{LS} I.M. Lifshitz and V. V. Slyozov, J. Phys. Chem. Solids \textbf{19}, 35
(1961).
\bibitem{W} C. Wagner, Z. Elektrochem. \textbf{65}, 581 (1961).
\bibitem{MS} B. Meerson and P.V. Sasorov, Phys. Rev. E \textbf{53}, 3491 (1996).
\bibitem{CP} J. Carr and O. Penrose, Physica D \textbf{124}, 166 (1998).
\bibitem{GMS} B. Giron, B. Meerson, and P.V. Sasorov, Phys. Rev. E \textbf{58}, 4213
(1998).
\bibitem{Pego} B. Niethammer and R. Pego, J. Stat. Phys. \textbf{95}, 867 (1999).
\bibitem{CMPS} M. Conti, B. Meerson, A. Peleg, and P.V. Sasorov, Phys. Rev. E \textbf{65}, 046117
(2002).

\end{thebibliography}
\end{document}